\documentclass{svjour3}                     % onecolumn (standard format)

\RequirePackage{fix-cm}

\smartqed  % flush right qed marks, e.g. at end of proof

\usepackage{amsmath}
\usepackage{amsfonts}
\usepackage{undertilde}
\usepackage{graphicx}
\usepackage{epstopdf}
\usepackage{natbib}
\usepackage{subcaption}
\usepackage{multirow}
\begin{document}

\title{Bayesian analysis of Jolly-Seber type models}
\subtitle{incorporating heterogeneity in arrival and departure}


\author{Eleni Matechou         \and
        Geoff Nicholls         \and \\ %etc.
        Byron J. T. Morgan      \and %etc.
        Jaime A. Collazo         \and %etc.
        James E. Lyons
}


\institute{E. Matechou \at
              School of Mathematics, Statistics and Actuarial Science, University of Kent, Canterbury, UK \\
              Tel.: +44 (0) 1227 8 16036\\
              Fax: +44(0)1227 8 27932\\
              \email{e.matechou@kent.ac.uk}           %  \\
           \and
           G. Nicholls \at
            Department of Statistics, Oxford University, Oxford, UK
           \and
            B. J. T. Morgan \at
 School of Mathematics, Statistics and Actuarial Science, University of Kent, Canterbury, UK \\
           \and
           J. A. Collazo \at
U.S. Geological Survey, North Carolina Cooperative Fish and Wildlife Research  Unit,\\ Department of Applied Ecology, North Carolina State University, Raleigh, USA.
           \and
           J. E. Lyons \at
 U.S. Fish and Wildlife Service, Division of Migratory Bird Management,\\ Patuxent Wildlife Research Center, Laurel, MD 20708, USA.
}

\date{{\it Received October} 2007. {\it Revised February} 2008.  {\it
Accepted March} 2008.}

\date{Received: date / Accepted: date}

\maketitle

\begin{abstract}

We propose the use of finite mixtures of continuous distributions in modelling the process by which new individuals, that arrive in groups, become part of a wildlife population. We demonstrate this approach using a data set of migrating semipalmated sandpipers (\textit{Calidris pussila}) for which we extend existing stopover models to allow for individuals to have different behaviour in terms of their stopover duration at the site.
We demonstrate the use of reversible jump MCMC methods to derive posterior distributions for the model parameters and the models, simultaneously. The algorithm moves between models with different numbers of arrival groups as well as between models with different numbers of behavioural groups. The approach is shown to provide new ecological insights about the stopover behaviour of semipalmated sandpipers but is generally applicable to any population in which animals arrive in groups and potentially exhibit heterogeneity in terms of one or more other processes.
\end{abstract}

\textit{This draft manuscript is distributed solely for the purpose of scientific peer review. Its
content is deliberative and predecisional, so it must not be disclosed or released by reviewers.
Because the manuscript has not yet been approved for publication by the U.S. Geological Survey
(USGS), it does not represent any official USGS finding or policy.}

\section{Introduction}
\label{section:intro}

Capture-recapture (CR) data, arising when individuals are captured, individually marked, and followed over time, are often collected from wildlife populations. In some cases, more than one type of sampling is used, as in capture-recapture-resight (CRR) data when individuals can be detected without necessarily being caught or in capture-recovery data when individuals can be detected dead. Different sampling schemes can also result in more than one data set being collected from the same population, and these are often analysed using an integrated approach \citep{Besbeasal2002}. 

Population ecology models are employed to analyse CR, CRR etc. data in order to estimate, among other things, the size of the population and the probabilities of survival of the individuals. These models need to account for the sampling scheme, imperfect detection and often also for potential heterogeneity between individuals. This can arise either in their characteristics, for example survival probabilities, or in their behaviour, that could for example affect their detection probability. 

Population ecology models are referred to as Jolly-Seber (JS) type \citep{Jolly1965, Seber1965}, if they model the process by which new individuals enter the population. An example of a JS type model is the \citet{SchwarzArnason1996} model, which uses the idea of a super-population of animals, $N$, and the entry parameters, $\beta_{b-1},\ b=1,\ldots,K$ to denote the proportion of $N$ that were new arrivals on sampling occasion $b$, where $K$ is the number of samples. For a discussion of alternative JS type model formulations see section 8.2.3 in \citet{McCreaMorgan2014}.

Individuals enter the population either through birth or immigration and they often do so in groups. For example, migrating birds arrive at stopover sites in flocks rather than individually while juveniles of a species can emerge in a synchronous manner. If the number of arrival or emergence groups is known, then finite mixture models of continuous distributions, such as the normal, can be used to model the process by which new individuals enter the population. See for example \citet{Matechoual2014} who modelled the emergence of butterfly broods using a mixture of two normal distributions.

However in many cases the number of arrival groups, and hence the number of mixture components, is unknown. Model selection criteria, such as the Akaike information criterion \citep[][AIC]{Akaike1973}, have doubtful validity for selecting between models with different mixture components
\citep[see][chapter 6]{McLachlanPeel2000}. \citet{Pledgeral2010} refer to a comment by \citet{BurnhamAnderson2002} who suggest that the parameter estimates have to be in the interior of the parameter space for AIC to be valid. \citet{Cubaynesal2012} report relatively low success rates of AIC, the Bayesian information criterion \citep[][BIC]{Schwarz1978} and the integrated classification criterion (ICL--BIC), which is similar to BIC but has an additional penalty
for fuzzy clustering \citep[]{Biernackial2000},  in selecting the true number of mixture components in CR data. Finally, parameter estimates can be sensitive to model choice \citep{Pledger2000} and choosing a single model for inference can be undesirable as well as difficult.

\citet{Arnoldetal2010} demonstrated the use of reversible jump \citep[RJ]{Green1995} MCMC in the case of finite mixture models that are used to account for heterogeneity in capture probabilities in closed populations. They referred to RJMCMC as a useful means of selecting between models with different numbers of mixture components, or obtaining model-averaged estimates of parameters. RJMCMC has also been used in the capture-recapture literature \citep[for example]{Brooksal2000, KingandBrooks2008, Kingal2010} as a method for selecting model covariates, assessing whether model parameters are constant over time or time-varying or to compare models which allow for heterogeneity between individuals using random effects to models which assume a homogeneous population.

In this paper we demonstrate the use of finite mixture models to describe the arrival pattern of migrating semipalmated sandpipers (\textit{Calidris pussila}) at a stopover site in terms of mixtures of continuous distributions, specifically the normal distribution. This approach provides a biologically meaningful interpretation of the results in which each mixture component is treated as a flock, so that flocks can be compared in terms of their relative sizes and mean arrival times. We use RJMCMC to obtain the posterior distribution of the number of arrival groups and a model-averaged estimate of the arrival pattern. 

The data set of semipalmated sandpipers was first analysed in \citet{Matechoual2013} (M13) who extended the stopover model of \citet{Pledgeral2009} by proposing integrated models for stopover data on birds that are marked, and therefore
individually identifiable, together with raw count data of unmarked birds.  They modelled the
probability that an individual present at the stopover site will remain until the next sampling
occasion, termed retention probability, as a function of calendar time and of the unknown time the
individual has already spent at the site, which they referred to as its ``age''. These stopover models provide estimates of the population size and indirect estimates of the total stopover duration. Stopover sites provide an essential opportunity for migrating birds to break their journey, rest
and refuel. It is important to assess the significance of a site, an attribute which is based on
the number of migrants that use it and the duration of their stopover, as this can aid in
formulating conservation strategies aimed at non-breeding habitat for migrant shorebirds
\citep{Brownal2001} and in measuring the effects of management treatments
\citep{NicholsandWilliams2006, Lyonsal2008}.

However, in M13 all birds are assumed to behave independently and identically in terms of
their stopover duration, a feature which is known to be untrue for many migratory species
\citep{AlerstamLindstrom1990,LyonsandHaig1995,Cristoletal1999,DinsmoreandCollazo2003,Rubolinietal2004,Bishopetal2006}.
In this paper we also use finite mixtures to allow for
different behavioural groups, defined by their retention probability and hence stopover duration at the site.  Our results agree
with life history strategies (e.g., mating system) that purport differential migration strategies
among sexes to maximize fitness \citep{Rubolinietal2004} and
show that there are at least two
behavioural groups of semipalmated sandpipers.

Hence, the work in this paper demonstrates the use of RJMCMC that moves between finite mixture models with different numbers of
homogeneous groups in two directions: arrival groups and behavioural groups. By using RJMCMC we are able to quantify the uncertainty arising
from the need to estimate the number of mixture components in each direction instead of relying on model selection
criteria to choose the ``best'' model. Additionally, the posterior distributions of model parameters, or functions
of them, can be naturally averaged across the different models, if appropriate and desirable.  
In section \ref{section:models} we give a brief introduction to finite mixtures models and the RJMCMC algorithm. We present the data set of semipalmated sandpipers and the results in section \ref{section:appl2}. The details of the RJMCMC algorithm specific to the application are given as Supplementary Material. 

We verified our formulae and code by comparing our results to those obtained from a very simple but reliable and independently
coded rejection algorithm, suitable for simple data sets only. We also fitted synthetic data of similar size to the real data set and
checked convergence of our algorithm to known parameter values.

\section{Mixture models and RJMCMC}
\label{section:models}

The data are represented in $\mathbf X$ with $\mathbf{X}_i$ the $i^{\mbox{th}}$ data vector. In generic mixture models, the model parameters are:

\begin{itemize}
\item[-] $G$, the number of mixture components, 
\item [-] $\pmb\pi=(\pi_1,\ldots,\pi_G)$, $\sum_{g=1}^G\pi_g=1$, the mixing proportions, 
\item [-] $\pmb\eta=(\pmb\eta_1,\ldots,\pmb\eta_G)$ with $\pmb\eta_g$ the collection of parameters of the $g$th mixture component, and
\item[-] $\pmb\psi$, a collection of parameters that are not part of the mixture components. 
\end{itemize}

\noindent We write $\pmb\theta=(G,\pmb\pi,\pmb\eta,\pmb\psi).$ We seek an expression for the posterior distribution, 

\[\mathcal{P}(\pmb \theta| \mathbf X) \propto \mathcal{P}(\mathbf X| \pmb \theta) \mathcal{P}(\pmb \theta)\]

\noindent of the parameters in $\pmb \theta$, allowing the number of mixture components, $G$, and hence the number of parameters in $\pmb\theta$ to be estimated, where
 
\[\mathcal{P}(\mathbf{X}|\pmb{\theta})=\prod_i\sum_{g=1}^G\pi_g\mathcal{P}(\mathbf{X}_i|g, \pmb\eta_g,\pmb\psi)\]

\noindent and $\mathcal{P}(\pmb{\theta})$ is the joint prior of the parameters.
 
\noindent We summarise $\mathcal{P}(\pmb \theta| \mathbf X)$ using a RJMCMC algorithm. This has two update types: one for updating parameters within models, $\pmb\pi,\pmb\eta,\pmb\psi$, and one for updating the number of mixture components, $G$.

We update within-model parameters $\pmb\eta$ and $\pmb\psi$ using a standard single-update Metropolis-Hastings random walk, described for example in \citet[][section 5.3.2]{Kingal2010}. Mixing proportions, $\pmb\pi$, are updated as follows: two groups are chosen at random, say $a$ and $b$, $\epsilon$ is defined as
$\epsilon = \gamma(\pi_a+\pi_b)$, where $\gamma\in(0,1)$ is fixed and chosen during tuning, $x$ is drawn from
Unif(-$\epsilon$, $\epsilon$) and $\pi'_a$ and $\pi'_b$ are calculated by $\pi'_a=\pi_a + x$ and $\pi'_b=\pi_b -
x$. If $\pi'_a, \pi'_b \geq 0$ and $\pi'_a \leq (\pi_a+\pi_b)$ the standard Metropolis-Hastings acceptance probability is calculated.

The number of mixture components, $G$, is updated using a RJMCMC move. The proposal transition probability to a model with $G'$ mixture components and $\pmb\theta'$ parameters from a model with $G$ components and $\pmb\theta$ parameters is denoted by $P_G(G'|G)$. 

Suppose that the proposed move is to a model with $G'=G+1$ groups. We allocate mass to this newly formed group by removing some mass from an existing group. Specifically, the proposed proportion of individuals in this new group, $\pi'_{G+1}$, is generated by choosing one of the
existing $G$ groups at random, say group $a$, with probability $1/G$, drawing $x$ from Unif(0,
$\pi_a$), setting $\pi'_{G+1}$ equal to $x$ and $\pi'_a$ equal to $\pi_a-x$.%, that is $\pmb\pi'=(\pi_1,\ldots,\pi_{a-1},\pi'_a=\pi_{a}-x,\ldots,\pi'_{G+1}=x)$.  
The parameters for this proposed group, $\pmb\eta'_{G+1}$, are generated from their corresponding prior, $\mathcal{P}(\pmb\eta'_{G+1})$. %, and $\pmb\eta'=(\pmb\eta_1,\ldots,\pmb\eta_G,\pmb\eta'_{G+1})$ while $\pmb\theta'=(G+1,\pmb\pi',\pmb\eta',\pmb\psi)$.

Suppose that the proposed move is to a model with $G'=G-1$ groups. We choose $a$ from $\text{Unif}\{1,\ldots,G\}$ and $b$ from $\text{Unif}\{1,\ldots,a-1,a+1,\ldots,G\}$. We remove group $a$ and allocate its mass to group $b$.%, that is we set $\pi'_b=\pi_a+\pi_b$ and hence $\pmb\pi'=(\pi_1,\ldots,\pi_{a-1},\pi_{a+1},\ldots,\pi_{b-1},\pi'_b,\pi_b,\ldots,\pi_{G})$ (or vice versa if $a>b$) and $\pmb\eta'=(\pmb\eta_1,\ldots,\pmb\eta_{a-1},\pmb\eta_{a+1},\ldots,\pmb\eta_G)$, while $\pmb\theta'=(G-1,\pmb\pi',\pmb\eta',\pmb\psi)$.

The acceptance probability for a model with $G'=G+1$ groups is given by 
\begin{gather}
\alpha(\pmb\theta,\pmb\theta')={\mbox{min}}\left(1,\frac{\mathcal{P}(\pmb \theta'| \mathbf X)P_G(G|G+1)\frac{1}{(G+1)}\frac{1}{G}}{\mathcal{P}(\pmb \theta| \mathbf X)P_G(G+1|G)\frac{1}{G}\frac{1}{\pi_a}\mathcal{P}(\pmb\eta'_{G+1})}\right).
\label{equation:RJprob}
\end{gather} 
The Jacobian term \citep[see][p. 165]{Kingal2010} required in forming equation (\ref{equation:RJprob}) is equal to 1 because $G'$ and $\pi'_a, \pi_{G+1}'$ are, respectively, linear functions of $G$ and $\pi_a$ and $\pmb\eta'_{G+1}$ are generated from their prior. 

The reverse move, to a model with $G-1$ groups, is fully defined given the above and is presented in detail for the example considered in this paper as Supplementary Material.

\cite{Arnoldetal2010} provide a detailed description of RJMCMC for closed population models that allow for heterogeneity in capture probability and we provide R \citep{R} code and details of the algorithm for analysing a data set from a closed population of cottontail rabbits, also presented by \cite{Arnoldetal2010} as Supplementary Material. For the application we present in this paper, our population is instead open and exhibits heterogeneity in both arrival and departure. We give details of the algorithm in this case as Supplementary material and we make R code available on request from the first author.

Checking convergence of the chain is not straightforward in RJMCMC and similarly, running multiple chains can be computationally very demanding. We recommend, and employ in the application considered here, examining trace plots for individual parameters, conditional on $G$, and using single-chain diagnostics, such as the Geweke convergence diagnostic
\citep{Geweke1992}, incorporated in the R \cite{R} package \textit{coda} \citep{coda} for parameters in $\pmb \psi$, to conclude convergence. 

\section{Application}
\label{section:appl2}

\subsection{Data and parameters}

The stopover site is formed by the wetlands at
the Tom Yawkey Wildlife Center in South Carolina where the study, which spans $T=38$ days, took place in spring of 2001. Samples are collected on $K=29$ of these
days and there are $11$ null occasions when no sampling takes place. There are two
types of sampling occasions: on capture occasions, birds can be
caught using mist nets and uniquely marked before being released; on resight
occasions, marked birds can be detected and an imperfect
count of unmarked birds is obtained. These raw counts of unmarked birds form vector $\mathbf y$ of
length $T$ with $T-K$ missing entries corresponding to occasions when counts were not obtained.

Each of the birds that visited the site during the study has its own capture-recapture-resight history
(CRRH) and we let $H$ denote the number of distinct observed CRRHs of the $D$ birds that were marked.
For CRRH $\mathbf{x}_{h}=(x_{h1},\ldots,x_{hT})$, shared by $n_h$
birds, with $x_{ht}\in\{0, 1, 2\}$, 2, 1, and 0 signifying that the $n_h$ individuals were resighted, caught or
missed, respectively, on occasion $t$. Any bird that was never caught has the trivial history ${\mathbf 0}$. All CRRHs have 11 missing entries. The $H$ CRRHs are summarised in matrix
$\mathbf X$ of dimension $H\times T$ and their frequencies are recorded in vector $\mathbf n$.

The $\mathbf{y}$--data, formed by the raw counts, and the $\mathbf{X}$--data, formed by the $H$
unique CRRHs together with their frequencies in $\mathbf n$, are the two data sets to be
analysed using the M13 proposed integrated model which has two parts: one that builds on the \cite{Pledgeral2009} model for the
$\mathbf{X}$--data and a binomial model for the $\mathbf{y}$--data. 

The model parameters are:
\begin{itemize}
\item[-] $N$: super-population size. The total number of birds that became available for
    capture-resight during the study without necessarily being detected.

\item[-] $M$: number of arrival groups.

\item[-] $w_m$, $\mu_m$ and $\sigma_m$, $m=1,\ldots,M$: respectively, population fractions, mean arrival
    times and standard deviations of arrival times of the M arrival groups,
    $\sum_{m=1}^Mw_m=1$. The population fraction that arrived between occasions $b-1$ and $b$ is the entry parameter $\beta_{b-1}$. In terms of the mixture components,
    \[\beta_{b-1}=\sum_{m=1}^Mw_m\left\{F_m(b)-F_m({b-1})\right\},\ b=2,\ldots,T-1,\]
     where
$F_m(b)=P(X\leq b)$ when $X\sim N(\mu_m,\sigma^2_m)$. The first and last intervals are treated
as open-ended with 
\[\beta_0=\sum_{m=1}^Mw_mF_m(1)\]
and 

\[\beta_{T-1}=1-\sum_{m=1}^Mw_mF_m(T-1), \forall m,\]

\noindent ensuring that the entry parameters sum to 1 i.e. $\sum_{b=1}^T\beta_{b-1}=1$.

Fig. 2 in the Supplementary Material demonstrates the modelling of the arrival process in terms of the
normal mixture components and the entry parameters $\beta_{b-1},\ b=1,\ldots,T$.

\item[-] $G$: number of behavioural groups. Individuals that belong to the same group have
    common baseline retention probability which can be different from the corresponding
    probability of the other $G-1$ groups.

\item[-] $\pi_g$, $g=1,\ldots,G$: The population fractions of the G behavioural groups, with
    $\sum_{g=1}^G\pi_g=1$.

\item[-] $\phi_{gta}$, $g=1,\ldots,G$, $t=1,\ldots,T-1$, $a=t-b+1$: retention probability. The
    probability that a bird that belongs to behavioural  group $g$, present at the site on
    occasion $t$ and of ``age'' $a$ will remain at the site until occasion $t+1$. As mentioned in section \ref{section:intro}, ``age'' is used to refer to the unknown time an
individual has already spent at the site.

    For the particular application, retention probabilities are modelled as additive in
    calendar time and ``age'', on the logit scale, with a different intercept for each group:
    
\[\mbox{logit}(\phi_{gta})=\gamma^{\phi}_{0g}+\gamma^{\phi}_1 t + \gamma^{\phi}_2 a,\]

\noindent where $\mbox{logit}^{-1}(\gamma^{\phi}_{0g})$ is the baseline retention probability for
    behavioural group $g$.

\item[-] $p_{t}$, $t=1,\ldots,T$: capture probability. The probability that a bird will be
    caught on occasion $t$ given that it is present.

 For the application considered in this paper, the number of nets used and the number of
hours they were left open on each capture occasion are multiplied together to form a covariate
for capture probability called ``effort'' ($e$) and two dummy variables, $loc2$ and $loc3$, are
created to model the effect of the three different locations where capture occasions took place
during the study, in an additive logistic regression model for capture probability: 

\[\mbox{logit}(p_{t})=\gamma^{p}_{0}+\gamma^{p}_1 e_t + \gamma^{p}_2 \mathcal{I}(loc2_t=1)  +
\gamma^{p}_3 \mathcal{I}(loc3_t=1),\]

\noindent where the indicator variable $\mathcal{I}(locj_t=1)$ is 1 if capture took place on location $j$, $j=2,3$, at time $t$ and 0 otherwise.

\item[-] $s_{t}$, $t=1,\ldots,T$: Resighting probability. The probability that a bird will be
    seen on occasion $t$ given that it is present. It is assumed to be the same for marked and
    unmarked birds and is modelled as constant, $s$, because resight occasions were
    conducted by the same crew which visited the same sites for the same length of time
    throughout the study.

\end{itemize}

The full set of parameters is

\[\pmb \theta=\left\{M, G,
(w_m, \mu_m, \sigma_m)_{m=1,\ldots, M},  (\pi_g, \gamma^{\phi}_{0g})_{ g=1,\ldots,G}, N,
\gamma^{\phi}_1, \gamma^{\phi}_2, \gamma^{p}_0, \gamma^{p}_1, \gamma^{p}_2, \gamma^{p}_3,
s\right\}.\]

In contrast to retention probabilities, dependence of capture and resight probabilities on
``age'' is not biologically meaningful and hence these parameters have not been modelled in terms
of $a$, but if necessary such a dependence can straightforwardly  be allowed for in the model.
Similarly, allowing for heterogeneous groups in terms of capture/resight probabilities in the
model is also possible in general, but it was not done here because of the small number of recaptures (5).

\subsection{Model, prior and posterior}

\subsubsection{Model}

Birds with CRRH $h$ have known times of first capture, $f_h$, and last detection, $l_h$, but unknown times of arrival, $b$, and departure, $d$. Let $\mathbf{z}=(g, b, d)$ denote the unknown life history of an individual. We can write 

\[\mathcal{P}(\mathbf{z}|\pmb \theta)=\pi_g\beta_{b-1}\left(\prod_{t=b}^{d-1}\phi_{gta}\right)
(1-\phi_{gda})^{{\mathcal I}(d<T)},\]

\noindent where the indicator variable ${{\mathcal I}(d<T)}$ is used to denote that the
departure of individuals still present at the end of the study cannot be observed.

If $\Omega_{\mathbf{z}} = \{(g, b, d): 1\leq b\leq f_h\leq l_h\leq d\leq T, g\in \{1,\ldots,G\}\}$ then the probability of CRRH
$\mathbf{x}_h$, $h\in \{1,\ldots,H\}$, given life history $\mathbf{z}$ and parameters in $\pmb \theta$
is
\[\mathcal{P}(\mathbf{x}_h|\mathbf{z}, \pmb \theta) = \left\{\begin{array}{l}
\left[\prod_{t=b}^{d}\left\{p_{t}^{{\mathcal I}(x_{ht}=1)}(1-p_{t})^{{\mathcal I}(x_{ht}=0)}\right\}^{{c_t}}\right]\times\\
\left[\prod_{t=f_{h}}^{d}\left\{s_{t}^{{\cal I}(x_{ht}=2)}(1-s_{t})^{{\cal I}(x_{ht}=0)}\right\}^{{r_t}}\right], \mathbf{z}\in \Omega_\mathbf{z}\\ \\ 0, \mbox{otherwise}\end{array}\right.\]

\noindent where variable $\mathcal{I(\psi)}$ is equal to 1 if condition $\psi$ is satisfied and 0
otherwise, $c_t=1$ if capture took place on occasion $t$ and variable $r_t=1$ if instead resighting
took place on occasion $t$, and 0 otherwise.

Similarly, if $\Omega'_{\mathbf{z}} = \{(g, b, d): 1\leq b\leq d\leq T, g\in \{1,\ldots,G\}\}$ the probability
of the ${\mathbf 0}$ history, given $\mathbf{z}$, $\pmb \theta$ is \[\mathcal{P}({\mathbf 0}|\mathbf{z}, \pmb \theta) = \left\{\begin{array}{l}
\left\{\prod_{t=b}^{d}\left(1-p_{t}\right)^{{c_t}}\right\}, \mathbf{z}\in \Omega'_{\mathbf{z}}\\ \\ 0, \mbox{otherwise}\end{array}\right..\]

Finally, 
\[\mathcal{P}(\mathbf X, \mathbf n|\pmb \theta) = \frac{N!}{\prod_h n_h!(N-D)!}\prod_h\left\{\sum_{\mathbf{z}\in\Omega_\mathbf{z}} \mathcal{P}(\mathbf{z}|\pmb \theta)\mathcal{P}(\mathbf{x}_h|\mathbf{z},\pmb \theta)\right\}^{n_h}\left\{\sum_{\mathbf{z}\in\Omega'_\mathbf{z}}\mathcal{P}(\mathbf{z}|\pmb \theta)\mathcal{P}({\mathbf 0}|\mathbf{z}, \pmb \theta)\right\}^{N-D}.\]

M13 treated the number of unmarked birds counted on resight occasion $t$, $y_t$, as a
binomially-distributed random variable with number of trials equal to $N$ and probability of
success the probability that a bird is present, unmarked and detected on that occasion. In this
case, the probability of success on occasion $t$ is 
\[\zeta_t=\sum_{g=1}^G\sum_{b=1}^t\pi_g\beta_{b-1}\left(\prod_{k=b}^{t-1}\phi_{gka}\right)
\left\{\prod_{k=b}^{t}(1-p_{k})^{c_k}\right\}s_{t}\]
and $y_t|\pmb \theta \sim \mbox{Bin}(N, \zeta_t)$. Therefore, $\mathcal{P}(\mathbf y|\pmb \theta) = \prod_{t=1}^T \mathcal{P}(y_t|\pmb\theta)^{r_t}$, where $r_t$ is as defined above.

\subsubsection{Prior}

Unless otherwise stated, simple, uninformative, independent priors were chosen for the model parameters. Specifically, we consider a Unif$\{1,\ldots,20\}$ prior for $M$ and a shifted Poisson with mean 1 for $G$, i.e. $G-1 \sim \mbox{Po}(1)$, as it is anticipated that there
    are mainly two types of stopover duration behaviours, namely the short and long type. For $N$ we take a N$(55000, 10000^2)$ prior with the mean chosen to be close to the point
    estimate obtained by M13, as this reflects our expectation for the size of the
    super-population. $w_m$, $m=1,\ldots,M$ and $\pi_g$, $g=1,\ldots,G$ are given Dirichlet priors with concentration parameters all equal to 1. For the logistic regression coefficients for $\phi$ and $p$ we followed \citet{Newman2003} and \citet[][p.
246]{Kingal2010} who suggested mean--0 normal priors with variances equal to
$\frac{\pi^2}{3(n+1)}$ where $n$ is the number of covariates in the model. Finally, the prior for $s$ is chosen as Beta(1, 1).

\subsubsection{Posterior}

Following M13, 
\[\mathcal{P}(\pmb \theta| \mathbf X, \mathbf n, \mathbf y)\propto \mathcal{P}(\mathbf X, \mathbf n| \pmb \theta)\mathcal{P}(\mathbf y|\pmb \theta) \mathcal{P}(\pmb \theta),\]
 where $\mathcal{P}(\pmb \theta)$ is the joint prior of the parameters in $\pmb \theta$.

\subsection{Results}

% All effective
%sample sizes were greater than 1000, including for parameters involved in the mixture components in
%models where the chain spent at least 5\% of the time. Alternative priors for $M$ and $G$ were also
%considered and although their resulting posterior distributions were different (for instance, when a
%uniform prior is chosen for $G$ the chain spends an almost equal amount of time in models with 2--4
%groups), the model-averaged estimates for the other parameters in $\pmb \theta$ are practically
%identical.

\begin{figure}[!h]
\centering
    \begin{subfigure}[]{0.4\textwidth}
            \includegraphics[scale=0.35]{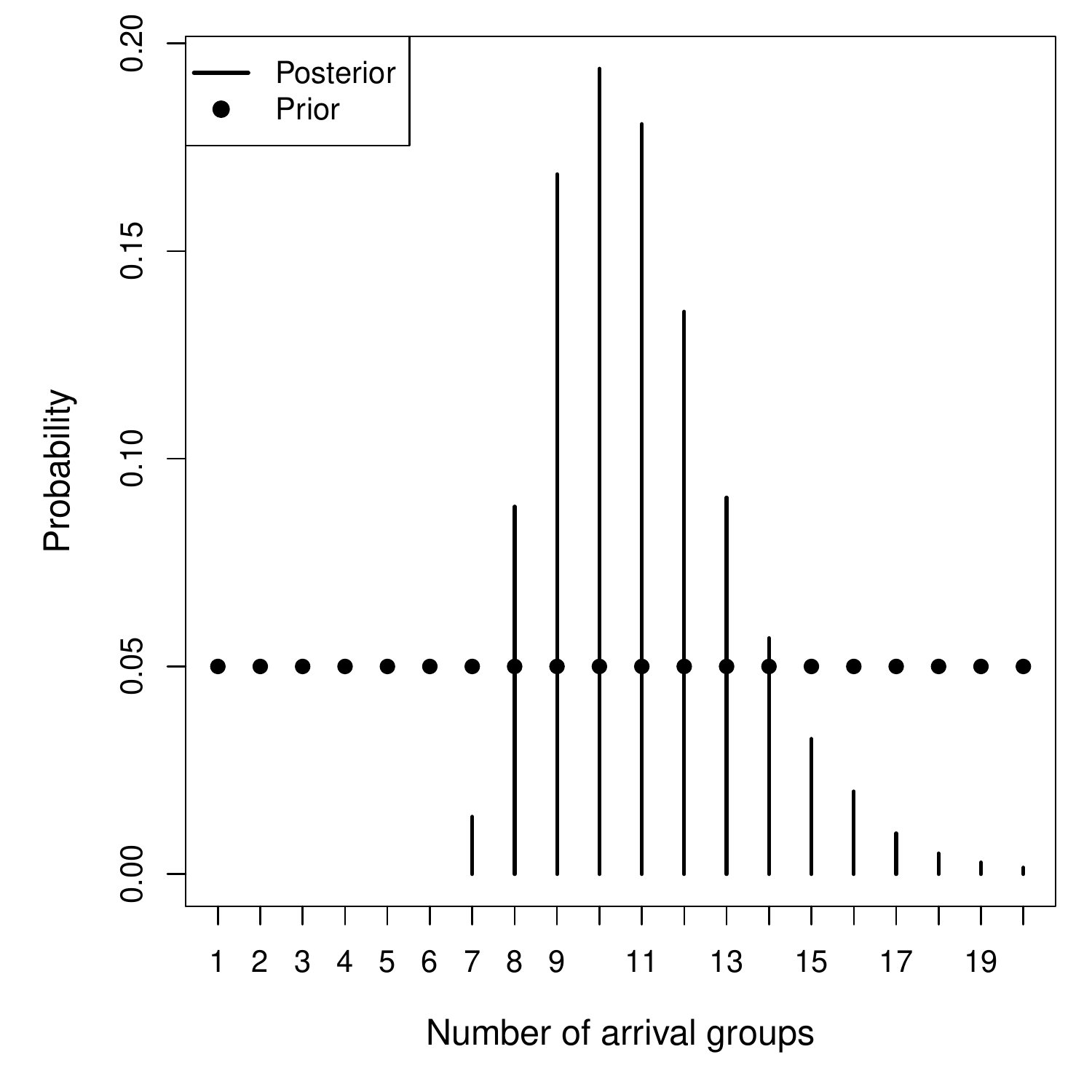}
                \caption{}
    \end{subfigure}
    \vspace{2cm}
    \begin{subfigure}[]{0.4\textwidth}
            \includegraphics[scale=0.35]{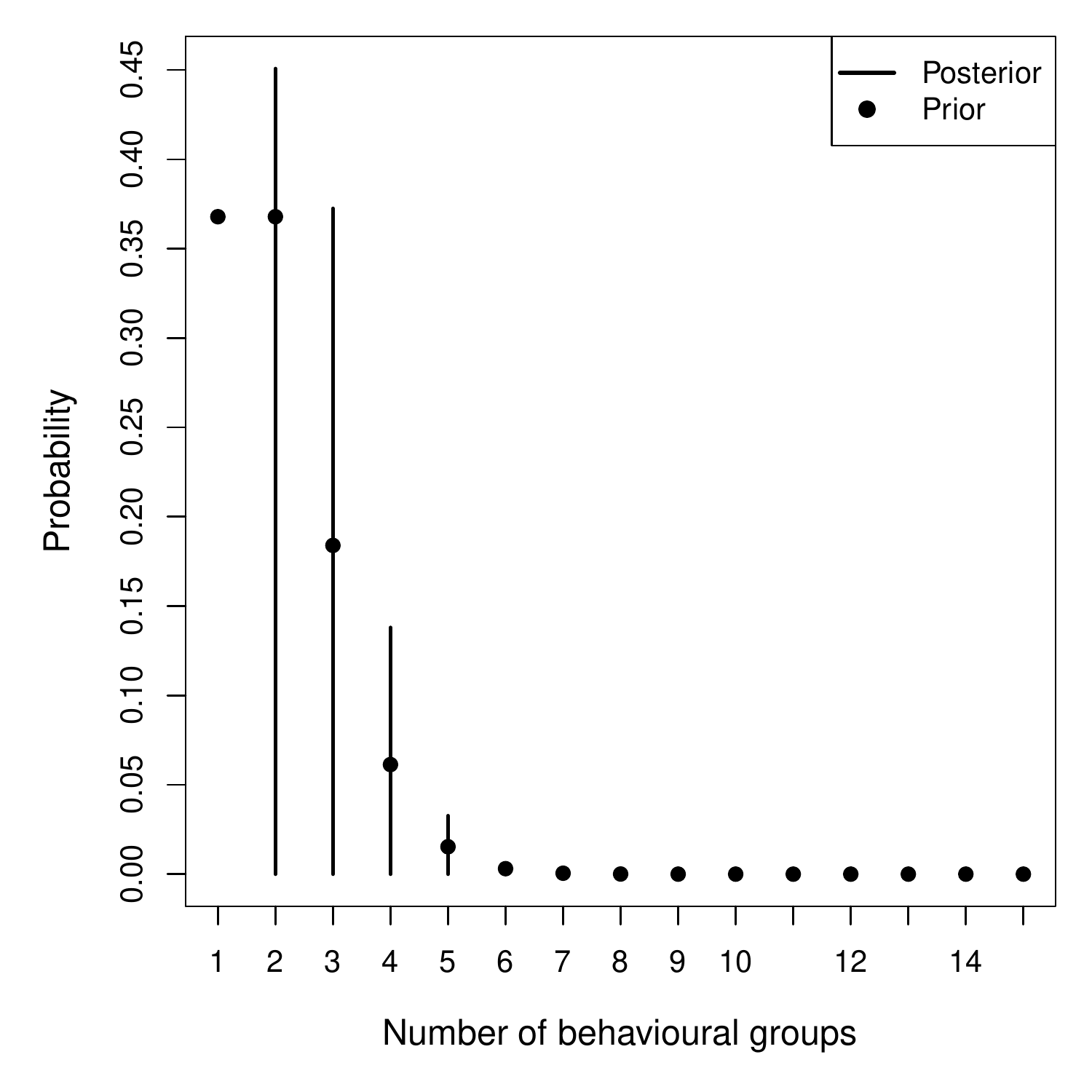}
                            \caption{}
    \end{subfigure}
    \vspace{-2cm}
    \caption{Prior and posterior distribution of, (a), $M$, the number of arrival groups and, (b), $G$, the number of behavioural groups.}\label{fig:MGpost}
\end{figure}

 The posterior distributions obtained for $M$ and $G$ are shown in Fig.
\ref{fig:MGpost}. The first peaks at $M=10$ and sharply declines for values of $M<9$, while its
right tail is longer. The chain spent over 90\% of its time in values of $M \in \{8,\ldots,13\}$.
The latter posterior peaks at $G=2$ and shows that the chain spent over 80\% of its time in models with $G=2$
or $G=3$. It gives no support to the model with $G=1$, suggesting that there are at least two
behavioural groups.

\begin{figure}[!h]
\centering

                \includegraphics[width=0.5\textwidth]{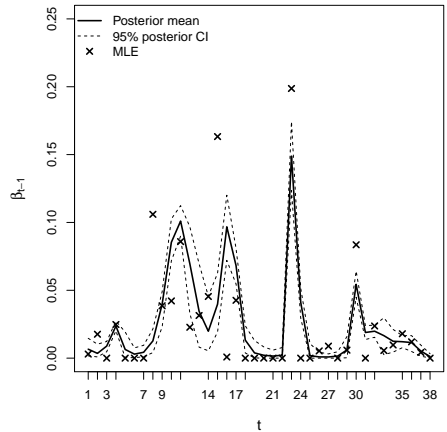}
                    %\vspace{-1cm}
                \caption{Model averaged posterior means and 95\% credible intervals for entry parameters, together with the maximum-likelihood estimates obtained by M13. The tick marks on the x-axis indicate days when sampling took place.}
                \label{fig:beta}
\end{figure}

Because the values for $M$ in $\{9,\ldots,12\}$ are almost equally well supported, with $M=8$ and $M=13$ also visited quite frequently by the chain, there is no clear choice for a best, or even top 2-3 models for $M$. Therefore, we present the model-averaged posterior distribution obtained for the entry parameters, $\beta_{t-1},\ t =
1\,\ldots,T$ in Fig. \ref{fig:beta}. For comparison, we also plot the maximum-likelihood estimates
obtained by M13, who estimated one entry parameter for each sample to model the arrival process. The latter are, mostly, included in the 95\% posterior credible intervals, with
the exception of the point estimates obtained corresponding to the modes of the four largest peaks, which are
all above the corresponding 97.5\% quantiles. Therefore, even though these two sets of estimates are not directly
comparable, since a different model for the $\beta$ parameters was used, they are very similar. The
M13 estimates, that are only constrained to sum to 1, are more flexible but do not reflect model-averaging uncertainty. On the other hand, our model-averaged estimates provide a smoother representation of the arrival
pattern of the birds at the site and are more robust to extremes. Our results suggest that the early arrival groups have greater spread in
their arrival times, with arrival times overlapping between groups, while the later arrival groups
are further apart and more distinct, with a longer right tail of arrivals right at the very end of
the stopover period.

\begin{figure}[!h]
\centering
    \begin{subfigure}[]{0.3\textwidth}
            \includegraphics[scale=0.25]{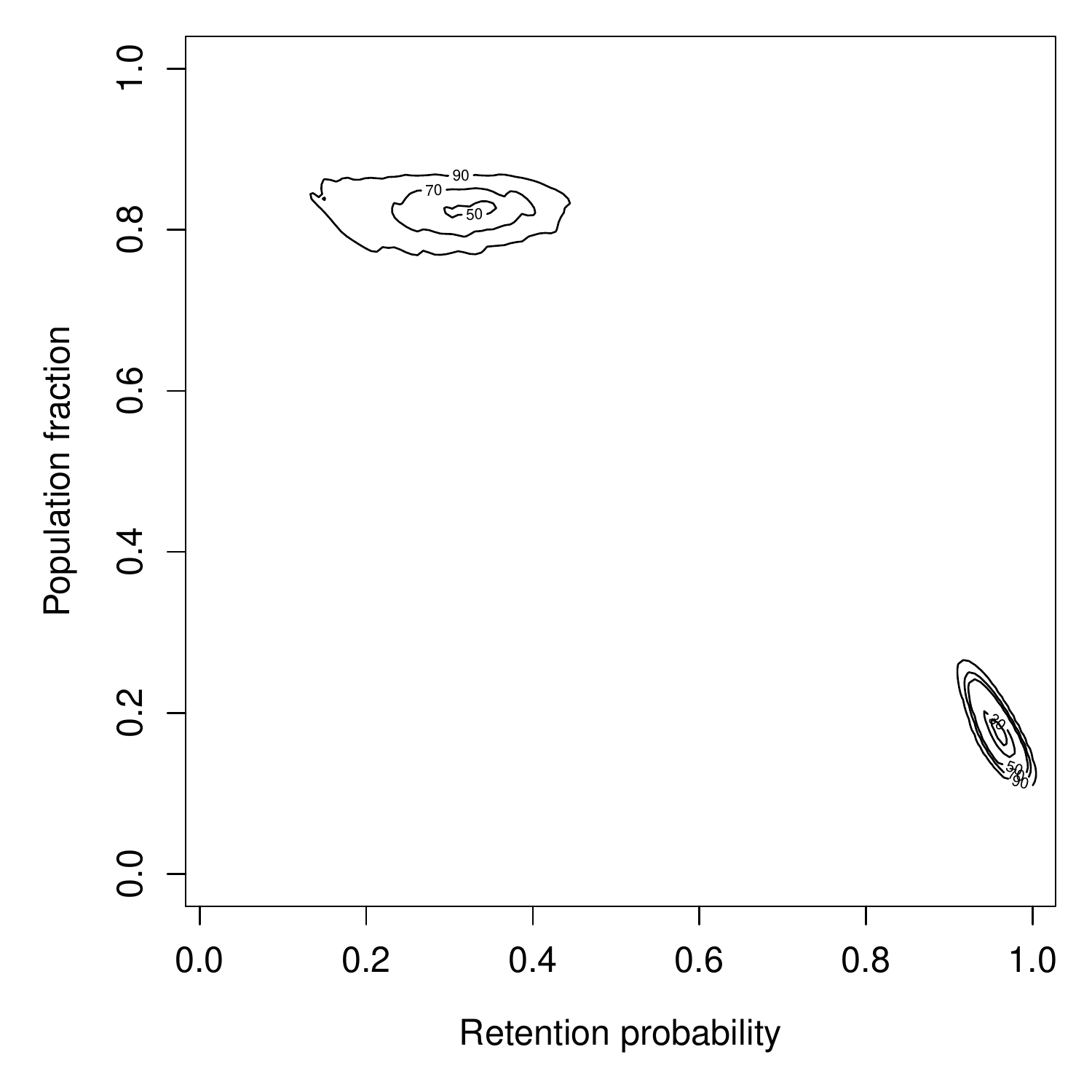}
                \caption{}
    \end{subfigure}
    %\hspace{0.7cm}
    \begin{subfigure}[]{0.3\textwidth}
            \includegraphics[scale=0.25]{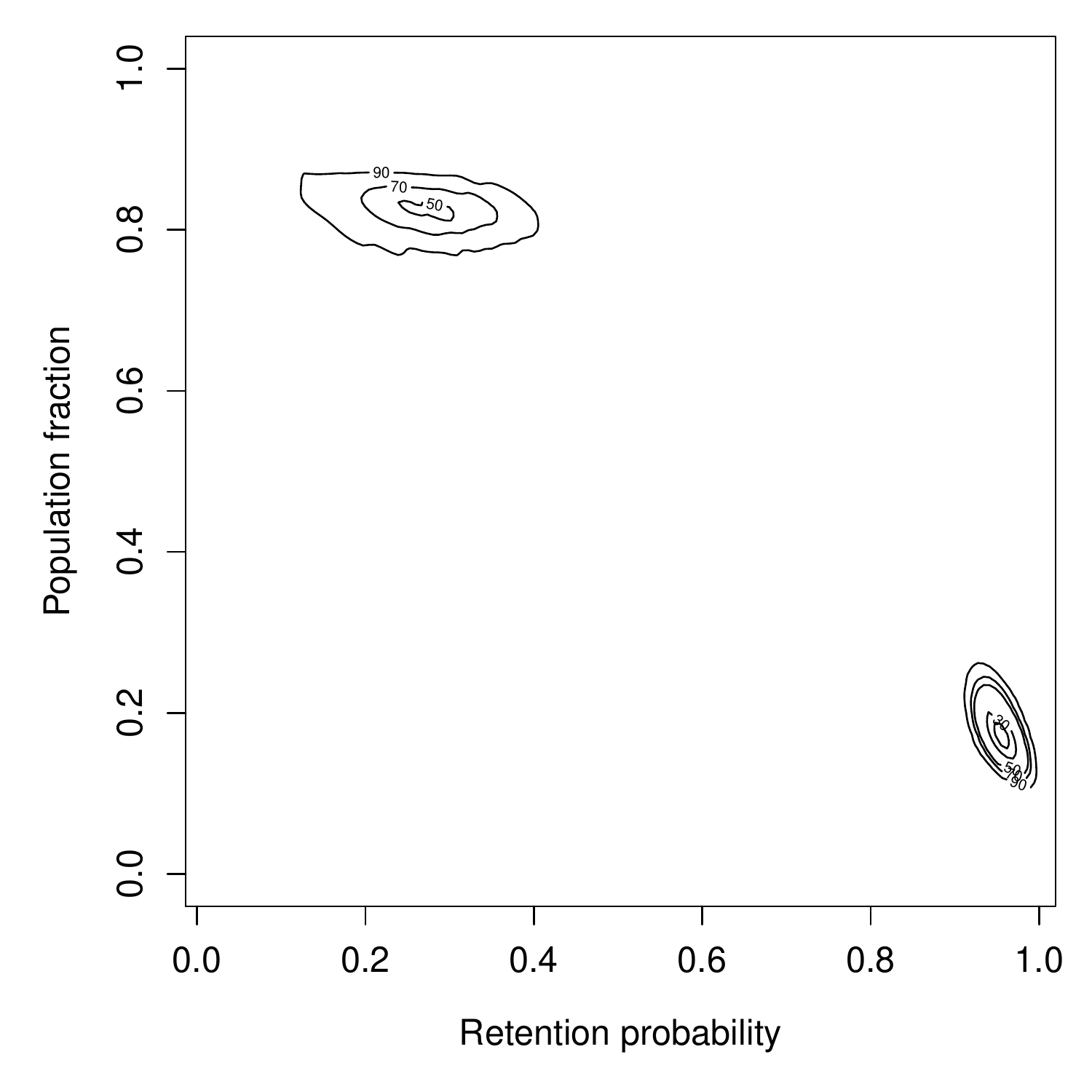}
                            \caption{}
    \end{subfigure}
    %\hspace{0.7cm}
    \begin{subfigure}[]{0.3\textwidth}
            \includegraphics[scale=0.25]{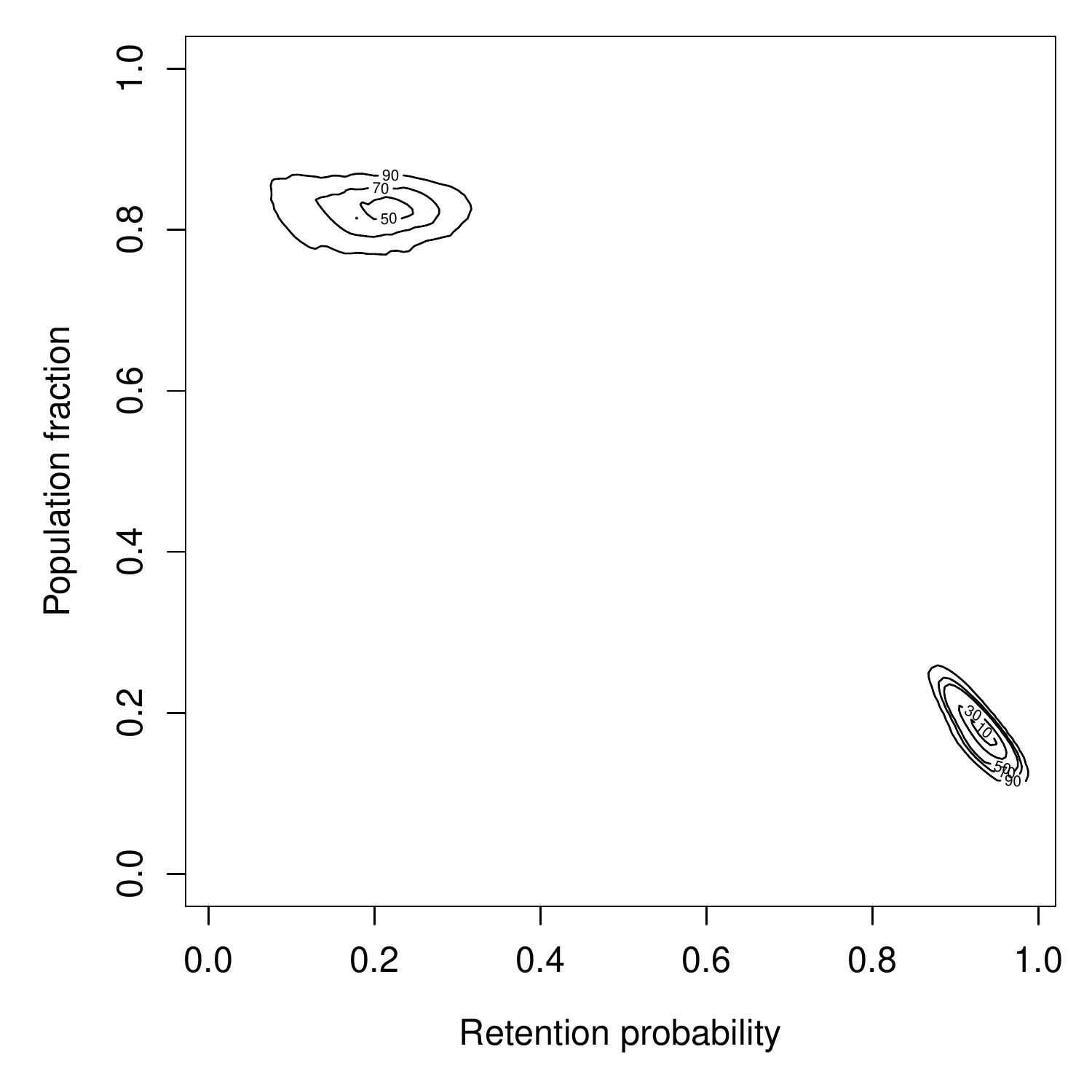}
                            \caption{}
    \end{subfigure}\\
    \begin{subfigure}[]{0.3\textwidth}
            \includegraphics[scale=0.25]{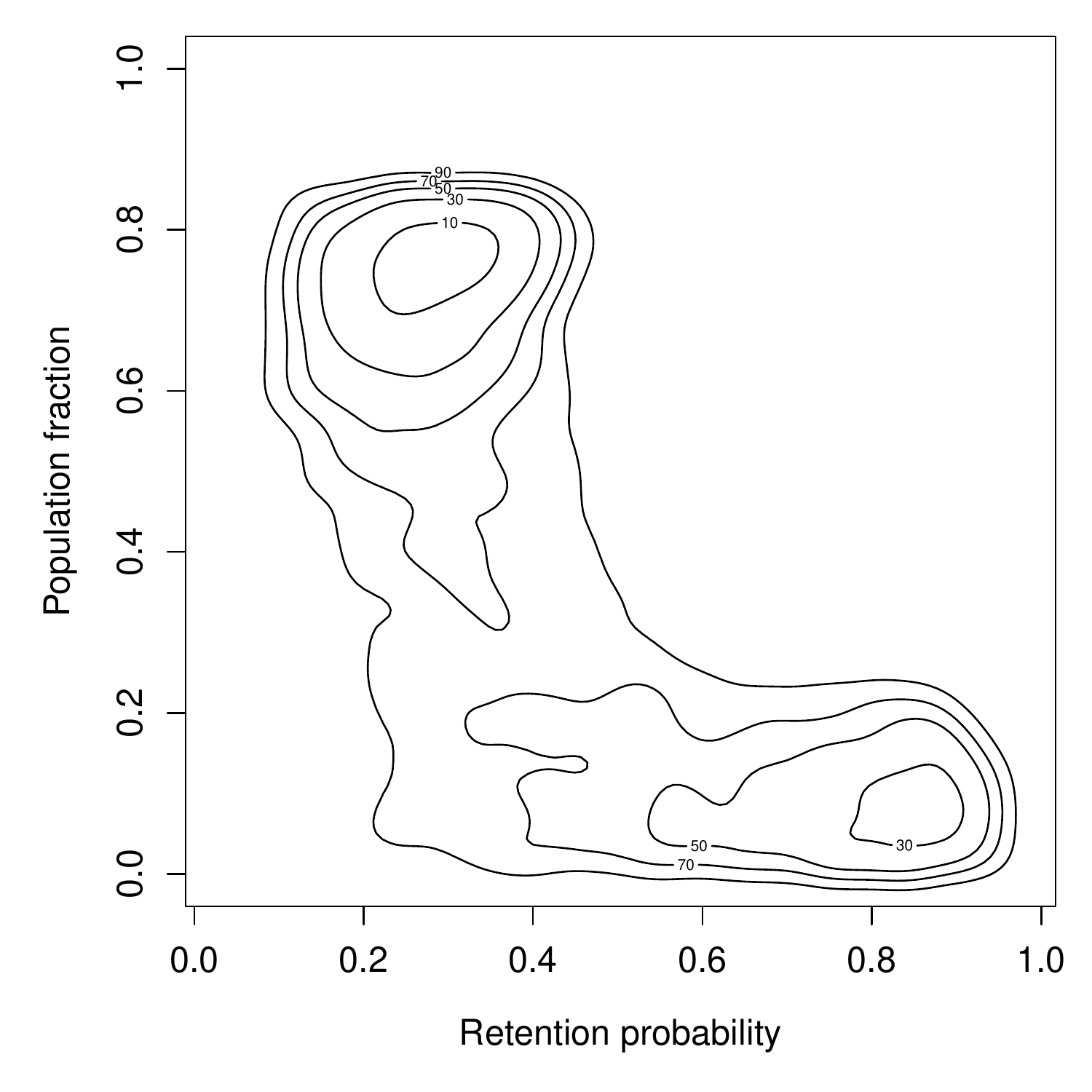}
                \caption{}
    \end{subfigure}
    %\hspace{0.7cm}
    \begin{subfigure}[]{0.3\textwidth}
            \includegraphics[scale=0.25]{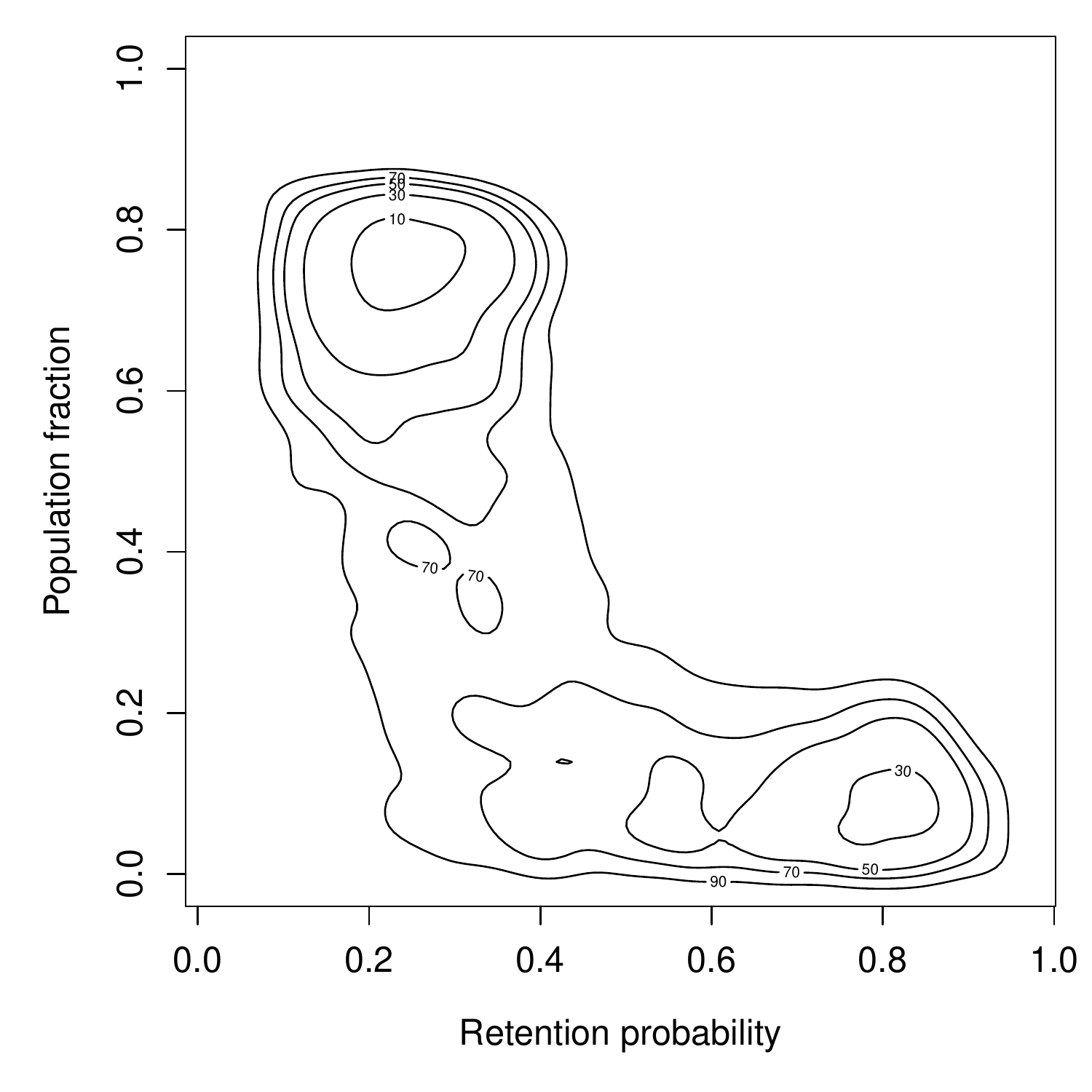}
                            \caption{}
    \end{subfigure}
    %\hspace{0.7cm}
    \begin{subfigure}[]{0.3\textwidth}
            \includegraphics[scale=0.25]{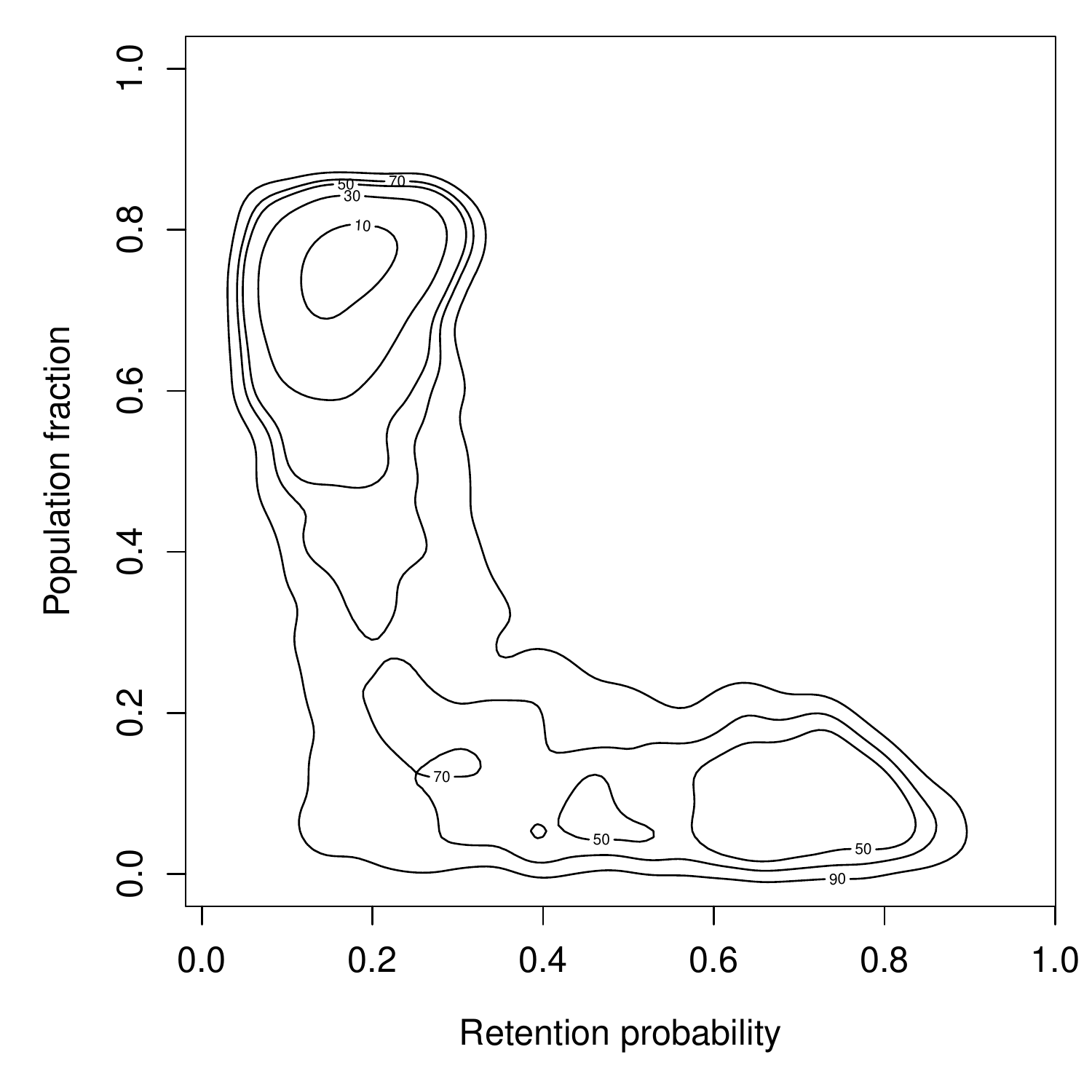}
                            \caption{}
    \end{subfigure}
    \caption{Contour plots of the joint posterior densities for retention probability and population fraction of each behavioural group when $G=2$ (top row) and $G=3$ (bottom row), when time $t=10$ and ``age'' $a=1$ (first column), time $t=10$ and ``age'' $a=10$ (middle column), time $t=20$ and ``age'' $a=1$ (last column).}
\label{fig:phi}
\end{figure}

The posterior densities of $\phi_{gta}$ and $\pi_g$ for $g=1,2$ when $G=2$ and $g=1,2,3$ when $G=3$ when $\{t=10,
a=1\}$, $\{t=10, a=10\}$ and $\{t=20, a=1\}$ are shown in Fig. \ref{fig:phi}. The areas of high
density suggest two very distinct groups: a large group, with population fraction $\approx$ 80\%
and low retention probability, and a small group, with population fraction $\approx$ 20\% with very
high retention probability. The areas of lower density when $G=3$ suggest that the third group, that connects the two
groups, has medium retention probability. These results are consistent with the predicted differences in migration strategies of male and female sandpipers; males may spend less time than females at stopover sites in order to reach the breeding sites earlier \citep{Bishopetal2004, Bishopetal2006}. Given the relative abundance of the retention groups in our study, this interpretation suggests a male-biased sex ratio in the stopover population. Alternatively, the heterogeneity in retention probability may reflect local movements of birds during a period of searching and settling that often occurs immediately after arrival to a stopover area \citep{AlerstamLindstrom1990}.  The group with low retention probability may be comprised of recent arrivals that were captured during a period of searching the landscape for favourable foraging conditions, but which ultimately settled outside the study area.  The smaller group with high retention probability may be comprised of birds that settled and remained in the study area during stopover.  Local movements may occur in response to changing conditions in prey abundance or water depth, facilitated by wetland connectivity \citep{FarmerandParent1997,Obernuefemannetal2013}. 

The effects of calendar time and ``age'' on retention probability are, as expected, found to be
negative, with model-averaged posterior means equal to -0.633 (95\% CI: -0.998, -0.346) and -0.145 (95\% CI:
-0.632, 0.310), respectively, although the effect of ``age'' is smaller than that of time with a
credible interval that includes 0. Since the logistic regression model used to model their effects
on retention probabilities is additive, plotting the joint distribution of retention probabilities
and population fraction of each behavioural group for different values of time and ``age'' simply
shifts the contours along the axis corresponding to retention probabilities, as the contour plots
in Fig. \ref{fig:phi} demonstrate.

In simulated data sets, obtained using parameter values in $\pmb \theta$ at 100 randomly chosen
simulation runs of the chain, conditional on $G=2$, the two  behavioural groups had an average
observed stopover duration in days equal to  1.3 (s.d. = 0.65)  and  9.95 (s.d = 7.12) while
conditional on $G=3$ the average observed stopover durations were  1.24 (s.d. = 0.58), 2.18 (s.d. =
2.25) and 10.33 (s.d. = 7.27) days.

The model-averaged mean of the posterior distribution of the super-population size, $N$, is equal to 61715 (95\%
CI: 53818, 71778), which is greater than the estimate of M13 but with overlapping confidence bands
(53595, asymptotic 95\% CI: 48349, 59410). The 95\% posterior credible intervals obtained for $N$ for all combinations of the most supported values for $M$ and $G$ are given in Table 1 in the Supplementary Material and agree with the model-averaged interval mentioned above.

To check the fit of our specified model, we considered parameter values in $\pmb \theta$ obtained
at 100 randomly chosen simulation runs of the chain. For each of these sets of parameter values, we
calculated the value of the log-likelihood for the real data set and for a data set simulated from
this set. The distributions of these two sets of log-likelihood values, plotted in Fig.
3 (a) in the Supplementary Material, peak at the same point, which suggests a good fit of the model. For each of
these simulated data sets we also obtained the number of birds first caught on each capture
occasion and the number of birds resighted as unmarked on each resighting occasion. Fig.
3 (b) and (c) in the Supplementary Material, respectively, show that the observed values almost always overlap with
the boxplots of these simulated values, once more providing support to the claim that the model
fits well.

\section{Discussion}
\label{section:discussion}

To analyse the stopover data set considered in section
\ref{section:appl2} we extended existing stopover models to allow for individuals to arrive in groups and to exhibit heterogeneity in their stopover duration at the site. We showed that both these processes can be modelled at the same time and the uncertainty in the number of groups in either process can be accounted for using a RJMCMC algorithm. Our results suggest that semipalmated sandpipers at stopover sites do not exhibit the same
behaviour in terms of their stopover. 

By using finite mixtures of continuous distributions,
such as normal, to model the arrival of individuals at the study site, instead of models with
fully time-dependent entry probabilities as in M13, the number of parameters does not necessarily increase
with the number of sampling occasions. Additionally, one is supplied with an uncomplicated and
biologically meaningful way to interpret the results in terms of the number of arrival groups and
their behaviour, making analyses on different data sets, for example from different years, directly
comparable. Additionally, the parameters of the mixture components, such as the mean arrival times,
can also be modelled as functions of, for example, weather covariates. Further simplifications of
the models are also possible. Specifically, it might be assumed that the means of the arrival times
of the different groups are equally spaced, with the space to be estimated by the model. We
have considered the case of normal mixtures for the work in this paper but other distributions, not
necessarily symmetric, could also be chosen, such as gamma, if appropriate.

Bayesian inference enables the incorporation of
prior beliefs which might not treat all models as equally likely a priori. The
belief that there are two behavioural group of birds at the stopover site, namely the short-- and the long--stayers was
easily incorporated in the model, instead of naively assuming that a model with 15
behavioural groups is as likely as the more realistic one with just two groups. On the other hand, posterior model probabilities are known to be sensitive to prior model probabilities \citep{CoraniandMignatti2015} and hence, the latter should be chosen very carefully.

The application presented demonstrates the general applicability of the RJMCMC algorithm, even when the population is heterogeneous in more than one processes, for instance both in survival and arrival and the models are highly complex. Unaccounted-for hererogeneity can lead to biased parameter estimates and spurious results. Specifically, it has been frequently
reported that unmodelled heterogeneity in capture probabilities leads to biased estimates of the
population size \citep{Pollockal1990}, but it can also affect estimation of survival
probabilities \citep{Oliveretal2011, Fletcheral2012,Matechoual2013b}. If potential heterogeneity in
survival probabilities remains unmodelled, then individuals with an overall higher survival probability will
prevail at older ages, which can result in the average survival probability appearing to
increase by age \citep{VaupelYashin1985, Peronal2010}, masking the effect of senescence. Accounting for heterogeneity is also important in non-ecological applications of CR models with an emphasis on estimating population size \citep[see][p. 46]{McCreaMorgan2014}.

Even when the list of possible models to be considered is large, as in the example of section \ref{section:appl2}, the use of the RJ algorithm makes model selection possible since inappropriate models do not have to be actually fitted, i.e. visited by the algorithm. When appropriate, the use of RJMCMC enables model-averaging and does not
require the quite often unclear or subjective choice of one single ``best'' model. The posterior density obtained for the number of arrival groups gives very similar support
to $M=9,\ldots,12$ groups, and the conclusions have been drawn by averaging over these, as
well as the less supported models.  

We have not considered heterogeneity in capture probabilities for the population of semipalmated sandpipers because of the low number of recaptures and we suggested in section \ref{section:appl2} that the models and the RJMCMC algorithm can be extended to that effect. However, it should be noted here that, as \citet{Link2003} explains and \citet{Arnoldetal2010} discuss, parameter $N$ is not identifiable among different model classes, for example between finite and infinite mixture models, and it may not be possible to distinguish between models with different assumptions about capture probability that provide very different estimates for $N$.

We demonstrated the models in this paper by considering a data set of migrating semipalmated sandpipers collected at a stopover site. However, the models are more generally applicable as other species and animals arrive or emerge in groups and exhibit heterogeneity in their survival or detection. For example they could apply to data sets of amphibians collected at breeding ponds, with different groups expected to arrive at different times \citep[observed male newts arriving before females]{Harrisonetal2009}, and in addition to have different detection and/or retention probabilities.

\begin{acknowledgements}
We are grateful to Fran\c{c}ois Caron for his suggestions and comments. Any use of trade, product,
or firms names is for descriptive purposes only and does not imply endorsement by the U.S.
government. The findings and conclusions in this article are those of the authors and do not necessarily represent the views of the U.S. Fish and Wildlife Service.
\end{acknowledgements}

\bibliographystyle{spbasic}\bibliography{bibliographia}

\clearpage

\section{Supplementary material}

\subsection{Demonstration of RJMCMC for the closed population of cottontail rabbits also considered in \cite{Arnoldetal2010}}

The population was closed, known to consist of $N=135$ uniquely identifiable individuals and sampled on $T=18$ occasions. $D=75$ rabbits were captured at least once.

Each of the $N$ individuals has its own capture history (CH) which is a sequence of T 1s and
0s, respectively corresponding to whether an individual was, or was
not, caught at each capture occasion. Let $H$ denote the number of distinct observed CHs of the $D$ individuals that were caught and let $h\in\{1,\ldots,H\}$ index these histories.
Additionally, let $\mathbf{x}_{h}=(x_{h1},\ldots,x_{hT})$ denote the CH shared by $n_h$ individuals. The $N-D$ individuals that were never caught have the trivial history ${\mathbf 0}$.  The $H$ distinct CHs are summarised in matrix
$\mathbf X$ of dimension $H\times T$ and their frequencies are recorded in vector $\mathbf n$.
%
%\citet{Pledger2000} found strong support for the model with two homogeneous groups. However, the maximum-likelihood estimate of population size was sensitive to the choice of the number of groups.

Suppose that there are $G$ different types or groups of rabbits, with individuals in the same group sharing the same capture probability, which varies between groups. The proportion of individuals in group $g$ is denoted by $\pi_g$, $g\in\{1,\ldots,G\}$ with $\sum_{g=1}^G\pi_g=1$. Consider a particular rabbit from group $g$. Denote by $p_{gt}$ the probability of detecting this rabbit on occasion $t$, $g=1,\ldots,G$ and $t=1,\ldots,T$.

The probability of CH
$\mathbf{x}_h$, $h\in \{1,\ldots,H\}$, given $g$ and parameters $\pmb \theta=(G, \pmb{\pi}, \pmb{p}, N)$
is $\mathcal{P}(\mathbf{x}_h|g, \pmb \theta) =
\prod_{t=1}^{T}\left\{p_{gt}^{x_{ht}}(1-p_{gt})^{1-x_{ht}}\right\}$. The probability an individual has the $\mathbf 0$ CH is $\mathcal{P}({\mathbf 0}|g, \pmb \theta) = \prod_{t=1}^{T}\left(1-p_{gt}\right)$.

Finally,
\[\mathcal{P}(\mathbf X, \mathbf n|\pmb \theta) = \frac{N!}{\prod_h n_h!(N-D)!}\prod_h\left\{\sum_{g=1}^G \pi_g\mathcal{P}(\mathbf{x}_h|g,\pmb \theta)\right\}^{n_h}\left\{\sum_{g=1}^G\pi_g\mathcal{P}({\mathbf 0}|g, \pmb \theta)\right\}^{N-D}.
\]

We take a prior non-informative with respect to scale, $1/N$, for the population size $N$. Our prior for the mixing proportions is uniform over all proportions that sum to 1. This choice is common in this context as it corresponds to a Dirichlet prior with concentration parameters all 
equal to 1. Capture probabilities depend on group but we assume that they are constant over time and we take uniform priors for the group capture probabilities $p_g$ themselves. Finally, we consider an uninformative Unif$\{1,\ldots,10\}$ prior for the number of capture groups.

Hence, $\mathcal{P}(\pmb \theta| \mathbf X, \mathbf n) \propto \mathcal{P}(\mathbf X, \mathbf n| \pmb \theta) \mathcal{P}(\pmb \theta)$, where $\mathcal{P}(\pmb \theta)$ is the joint prior of the parameters in $\pmb \theta$.

The algorithm has two update types; one for updating parameters within models, in this case parameters $\pmb\pi$, $\mathbf{p}$ and $N$, using standard Metropolis-Hastings steps, and one for updating $G$ using a RJ step. 

During the first type update, parameters $\pmb\pi$ are updated as follows: two groups are chosen at random, say $a$ and $b$, $\epsilon$ is defined as
$\epsilon = \gamma(\pi_a+\pi_b)$, where $\gamma\in(0,1)$ is fixed and chosen during tuning, $x$ is drawn from
Unif(-$\epsilon$, $\epsilon$) and $\pi'_a$ and $\pi'_b$ are calculated by $\pi'_a=\pi_a + x$ and $\pi'_b=\pi_b -
x$. If $\pi'_a, \pi'_b \geq 0$ and $\pi'_a \leq (\pi_a+\pi_b)$ the standard Metropolis-Hastings acceptance probability is calculated. $N$ is updated by proposing $N'$ from a Poisson distribution with mean $N$ with the acceptance probability calculated only if $N'>D$ and, finally,  $p_g$ is updated by proposing $p'_g$ from $N(p_g, \sigma^2)$, $g=1,\ldots, G$, with $\sigma^2$ chosen during tuning.

For the second type update, the proposal transition probabilities are $P_G(G+1|G) = P_G(G-1| G)
=0.5$, $G=2,\ldots,9$ and $P_G(2|1)=P_G(9 | 10)=1$.

If the proposed move is to a model with $G'=G+1$ groups then a value for the capture probability of the additional group is generated from the
uniform prior while for the proportion of individuals in this group a value is generated as described above.  The Unif(0,1) prior density is equal to 1 while the Dirichlet density with
all $G$ concentration parameters equal to 1 is equal to $(G-1)!$. Therefore, the prior for $(\pi_g, p_g)_{g=1,\ldots,G}$ is equal to
$G!(G-1)!$, with the $G!$ term accounting for
the fact that there are $G!$ indistinguishable ways to arrange the mixture components, each
resulting in the same representation of the data set. Hence the
acceptance probability from equation (1) in section 2 of the paper simplifies to:
\begin{equation*}
\begin{split}
\alpha(\pmb\theta,\pmb\theta') &  ={\mbox{min}}\left(1,\frac{\mathcal{P}(\mathbf X, \mathbf n| \pmb
\theta')G!(G+1)!P_{G}(G|G+1)\frac{1}{G+1}\frac{1}{G}}{\mathcal{P}(\mathbf
X, \mathbf n| \pmb \theta)(G-1)!G!P_G(G+1|G)\frac{1}{G}\frac{1}{\pi_a}}\right)\\
\smallskip
& = {\mbox{min}}\left(1,\frac{\mathcal{P}(\mathbf X, \mathbf n| \pmb \theta')GP_{G}(G|G+1)}{\mathcal{P}(\mathbf X, \mathbf n| \pmb \theta)P_G(G+1|G)\frac{1}{\pi_a}}\right),
\end{split}
\end{equation*}

\noindent since the priors for parameters that are common in the two models cancel out. The same holds for the prior for $G$, which is
symmetric and the priors for the new parameters, which cancel out with their corresponding proposal densities which appear in the denominator.

Similarly, the acceptance probability for a model with $G-1$ groups is:
\[\alpha(\pmb\theta,\pmb\theta')={\mbox{min}}\left(1,\frac{\mathcal{P}(\mathbf X, \mathbf n| \pmb
\theta')P_G(G|G-1)\frac{1}{\pi_a+\pi_b}}{\mathcal{P}(\mathbf X, \mathbf n| \pmb
\theta)(G-1)P_{G}(G-1|G)}\right).\]

\begin{figure}
\centering
    \begin{subfigure}[]{0.45\textwidth}
            \includegraphics[scale=0.35]{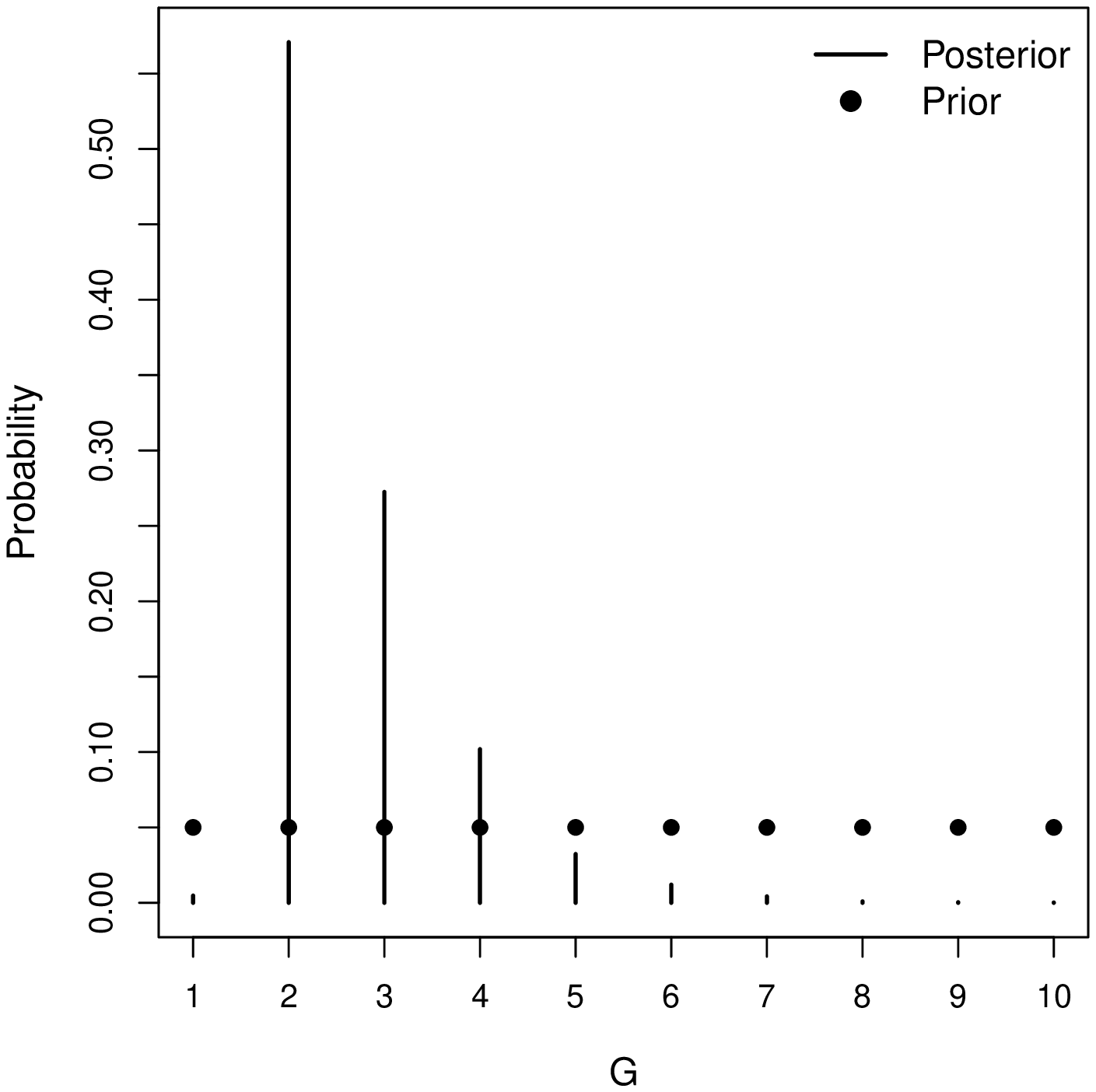}
                \caption{}
    \end{subfigure}
    \vspace{2cm}
    \begin{subfigure}[]{0.45\textwidth}
            \includegraphics[scale=0.35]{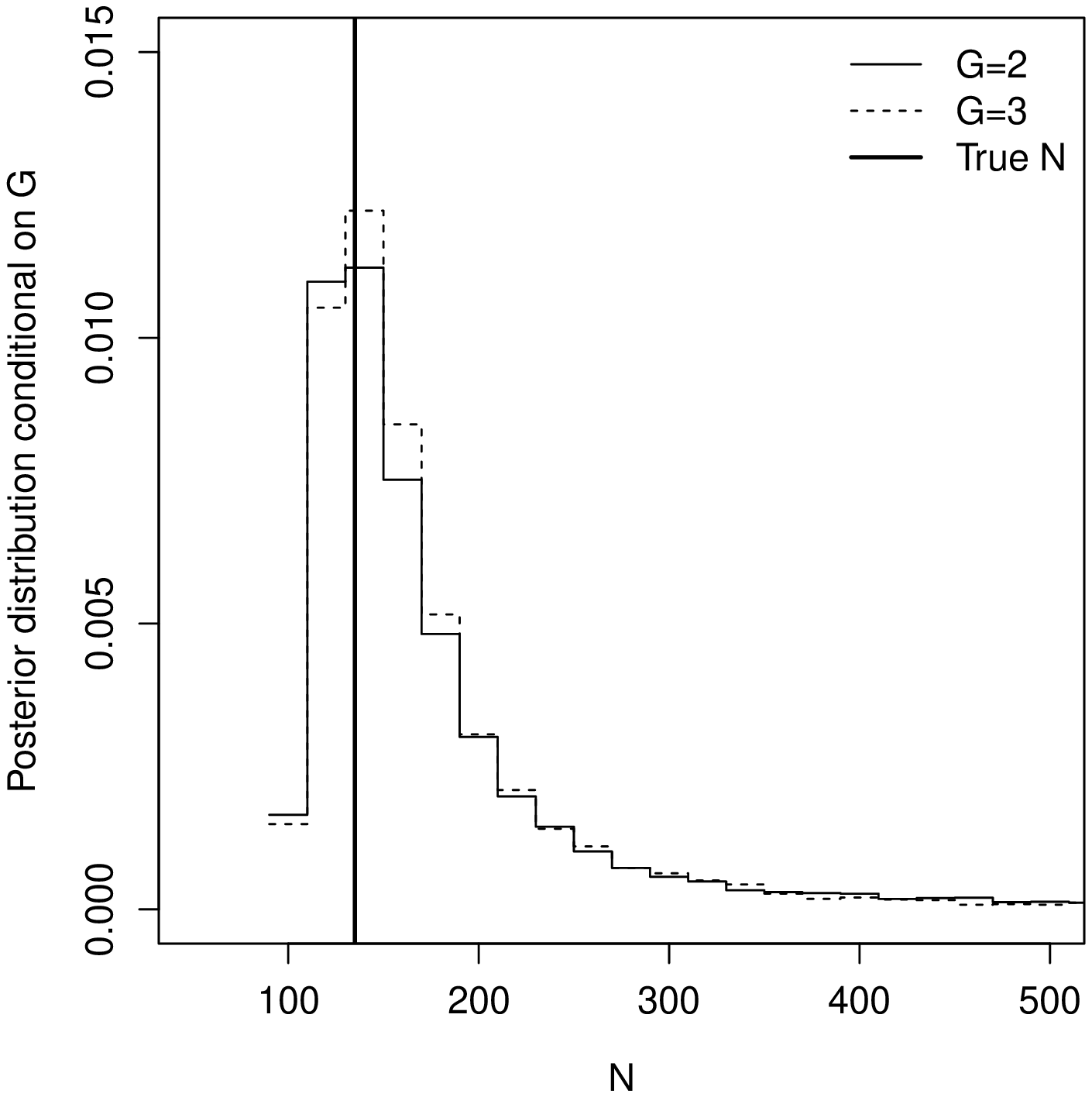}
               \caption{}
    \end{subfigure}\\
        %\vspace{0.3cm}
        \begin{subfigure}[]{0.45\textwidth}
            \includegraphics[scale=0.35]{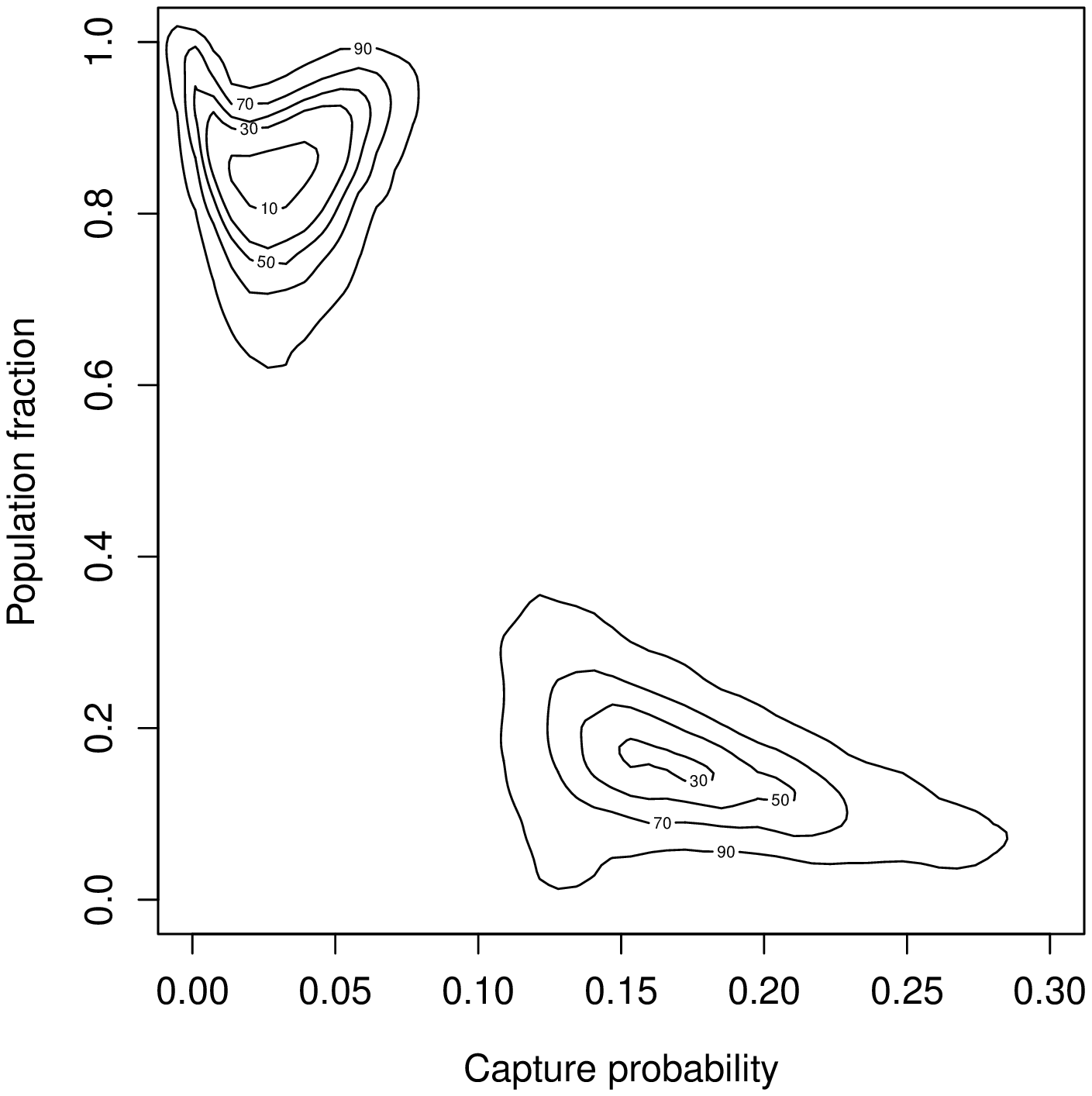}
              \caption{}
    \end{subfigure}
    \caption{(a): prior and posterior distribution of $G$, the number of capture groups. (b): posterior distribution of the population size $N$. The vertical line denotes the known true value. (c): contour plots of the joint posterior density of capture probabilities of the two groups, $p_1$ and $p_2$ and the proportions in each of the groups, $\pi_1$, $\pi_2$, conditional on $G=2$.}\label{fig:GNppipost_rab}
\end{figure}

The posterior distribution for the number of capture groups (Fig. \ref{fig:GNppipost_rab} (a)) peaks at $G=2$ and sharply declines for values of $G>3$, while it has very small mass at $G=1$, suggesting a heterogeneous population. 
The posterior distribution for the population size has a very long right tail, which has been truncated here in order for the area around the mode to be clearly visible. The median, which is equal to 143 when $G=2$ and 142 when $G=3$, is close to the true value (135) (Fig. \ref{fig:GNppipost_rab} (b)). Finally, both groups are found to have low capture probabilities, with the majority of rabbits having an estimated capture probability below 10\% (Fig. \ref{fig:GNppipost_rab} (c)). The contours are drawn at the shown percentages using function \texttt{kde} from the \texttt{ks} R \citep{R} package.

\clearpage

\subsection{Fig. 2, mentioned in section 3.1: entry parameters in terms of mixtures of $M=3$ normal distributions}

\begin{figure}[!h]
\centering
            \includegraphics[scale=0.5]{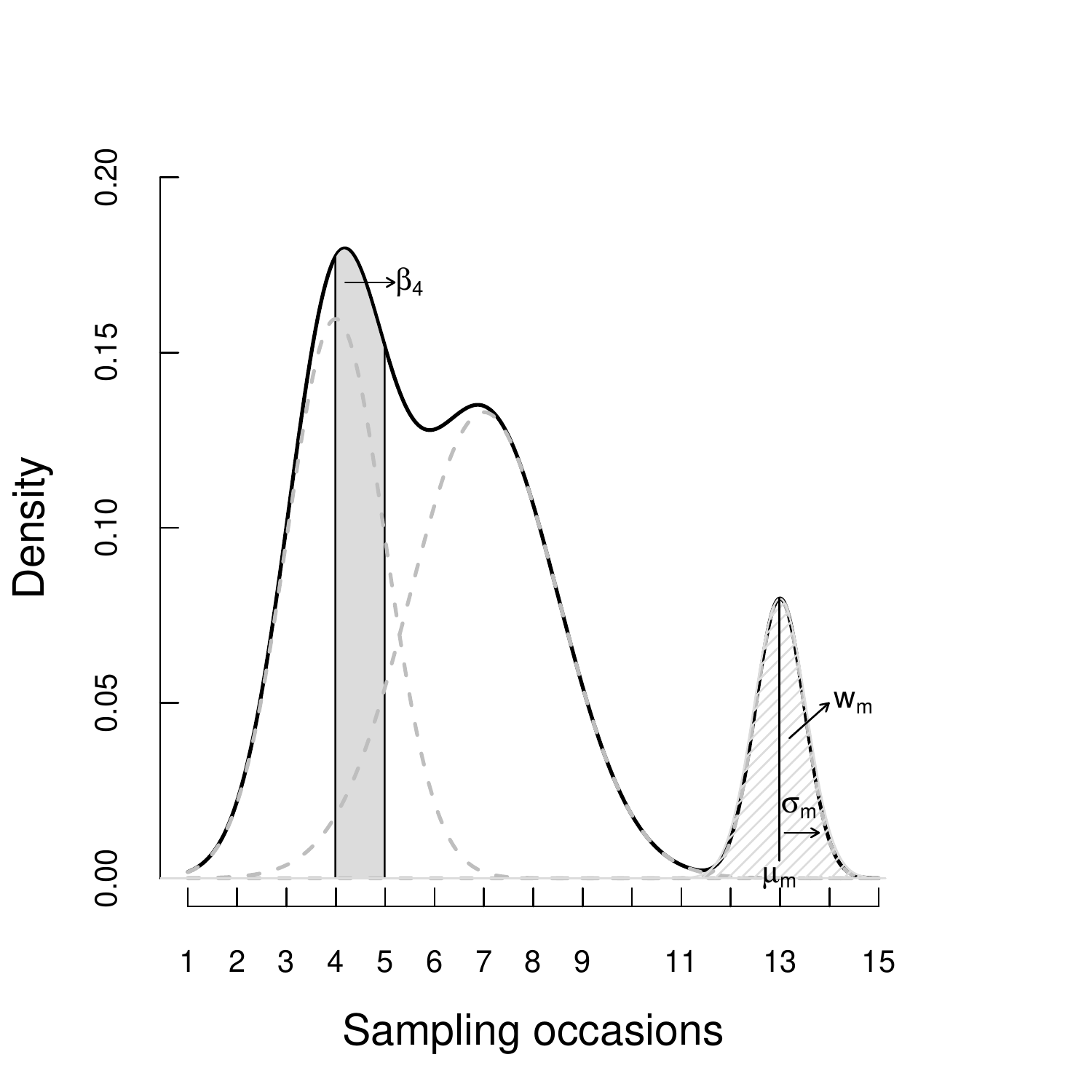}
    \caption{Demonstration of the modelling of the entry parameters in terms of mixtures of $M=3$ normal distributions. The black solid line represents the mixture density of the three individual mixture components with
    population fractions 0.4, 0.5, 0.1 and the grey dashed lines represent the scaled densities. The shaded grey area shows the resulting $\beta_4$ entry parameter.}\label{fig:example}
\end{figure}

\clearpage

\subsection{Details on the RJMCMC algorithm for the data set of semipalmated sandpipers presented in section 3}

During the first update type of the algorithm we use a single-update Metropolis-Hastings random walk for all parameters, with the exception of the proportions in each arrival and each behavioural group, $w_m, \
m=1,\ldots, M$ and $\pi_g, \ g=1,\ldots, G$, for which the updates are as described in section 2 of the paper.

We set $P_M(M+1|M) = P_M(M-1| M) = P_G(G+1|G) = P_G(G-1| G)
=0.5$, $M=2,\ldots,19$, $G=2,\ldots,14$ and $P_M(2|1)=P_M(19 | 20)=P_G(2|1)=P_G(14 |
15)=1$.

If the proposed move is to a model with $M'=M+1$ arrival groups then values for the mean, $\mu'$, and
standard deviation, $\sigma'$, of the arrival times of the additional group are generated from the
corresponding priors. For the proportion of individuals in this new group, $w'$, a value is generated as described in the previous section.

The priors for the parameters not involved in the normal mixture components are common between models with
different numbers of arrival groups, and therefore cancel  from the acceptance ratio as they
appear in both the numerator and denominator. The same holds for the prior for $M$, which is
symmetric.  The prior for $(w_m, \mu_m,\sigma_m)_{m=1,\ldots,M}$ is equal to
$M!(M-1)!\left\{\mathcal{P}_{\mu}(\mu)\mathcal{P}_{\sigma}(\sigma)\right\}^M$. Therefore, the acceptance probability for a model with $M+1$ arrival groups is given by: 

\begin{equation*}
\begin{split}
\alpha(\pmb\theta,\pmb\theta') &  ={\mbox{min}}\left(1,\frac{\mathcal{P}(\mathbf X, \mathbf n| \pmb
\theta')\mathcal{P}(\mathbf y|\pmb
\theta')M!(M+1)!\left\{\mathcal{P}_{\mu}(\mu)\mathcal{P}_{\sigma}(\sigma)\right\}^{M+1}P_{M}(M|M+1)\frac{1}{M+1}\frac{1}{M}}{\mathcal{P}(\mathbf
X, \mathbf n| \pmb \theta)\mathcal{P}(\mathbf y|\pmb
\theta)(M-1)!M!\left\{\mathcal{P}_{\mu}(\mu)\mathcal{P}_{\sigma}(\sigma)\right\}^{M}P_M(M+1|M)\frac{1}{M}\frac{1}{w_a}\mathcal{P}_{\mu}(\mu)\mathcal{P}_{\sigma}(\sigma)}\right)\\
\smallskip
& = {\mbox{min}}\left(1,\frac{\mathcal{P}(\mathbf X, \mathbf n| \pmb \theta')\mathcal{P}(\mathbf y|\pmb \theta')MP_{M}(M|M+1)}{\mathcal{P}(\mathbf X, \mathbf n| \pmb \theta)\mathcal{P}(\mathbf y|\pmb \theta)P_M(M+1|M)\frac{1}{w_a}}\right).
\end{split}
\end{equation*}

Similarly, the acceptance probability for a model with $M-1$ arrival groups is:

\[\alpha(\pmb\theta,\pmb\theta')={\mbox{min}}\left(1,\frac{\mathcal{P}(\mathbf X, \mathbf n| \pmb
\theta')\mathcal{P}(\mathbf y|\pmb \theta')P_M(M|M-1)\frac{1}{w_a+w_b}}{\mathcal{P}(\mathbf X, \mathbf n| \pmb
\theta)\mathcal{P}(\mathbf y|\pmb \theta)(M-1)P_{M}(M-1|M)}\right).\]

Updates for $G$ are performed in the same way. If the addition of a behavioural group is proposed,
the value for its intercept for retention probabilities is generated from the prior and for its
population fraction in the way described in the previous section. The
acceptance probabilities for moves between models with different numbers of behavioural groups are
therefore calculated in the same way, with $M$, $P_M$, $w_a$ and $w_b$ replaced by $G$, $P_G$,
$\pi_a$ and $\pi_b$, respectively. The only difference is that the priors for $G$ do not cancel
in this case and so the numerator and denominator are each multiplied by the corresponding shifted
Poisson probability.

\clearpage

\subsection{Table 1, mentioned in section 3.3: Credible intervals for $N$ for different values of $G$ and $M$}

\begin{table}[!h]
\centering
\caption{95\% posterior credible intervals for $N$ for all the combinations of the values for $M$ and $G$ with the highest posterior densities.}
\label{table:Nall}
\begin{tabular}{ccccc}
&\multicolumn{4}{c}{$G$}\\ 
$M$&\multicolumn{2}{c}{2}&\multicolumn{2}{c}{3}\\ \cline{2-5}
8&(54017,&69135)&(53861,&68610)\\ \hline
9&(53864,&71033)&(53398,&69337)\\ \hline
10&(53898,&71661)&(53517,&70472)\\ \hline
11&(55096,&73551)&(53833,&70418)\\ \hline
12&(54409,&74660)&(53765,&72720)\\ \hline
13&(55549,&73832)&(55154,&72781)\\ \hline
\end{tabular}
\end{table}

\clearpage

\subsection{Fig. 3 mentioned in section 3.3: GOF test for the sandpiper data set}

\begin{figure}[!h]
\centering
    \begin{subfigure}[]{0.45\textwidth}
            \includegraphics[scale=0.35]{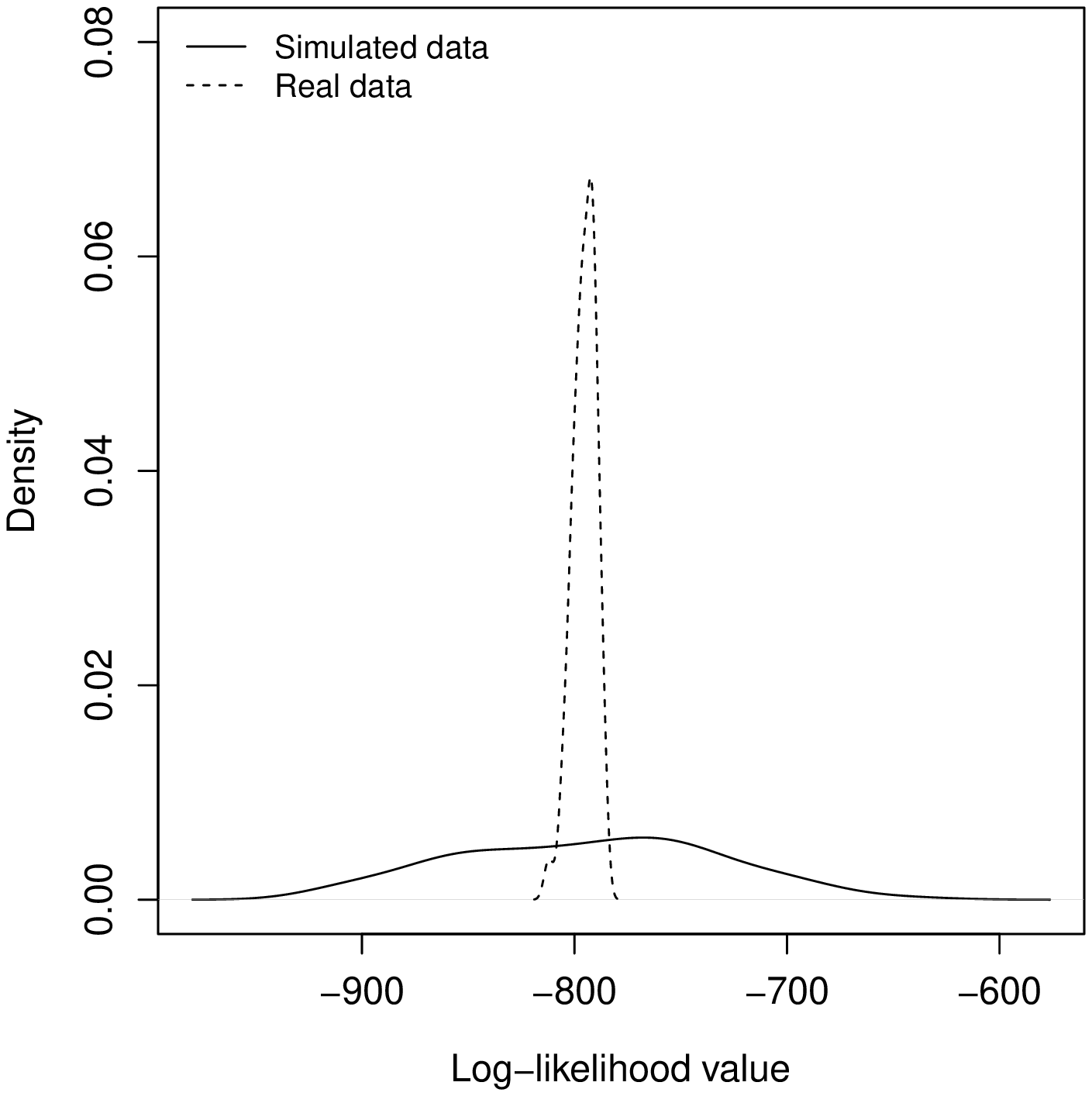}
                \caption{}
    \end{subfigure}
    \vspace{2cm}
    \begin{subfigure}[]{0.45\textwidth}
            \includegraphics[scale=0.35]{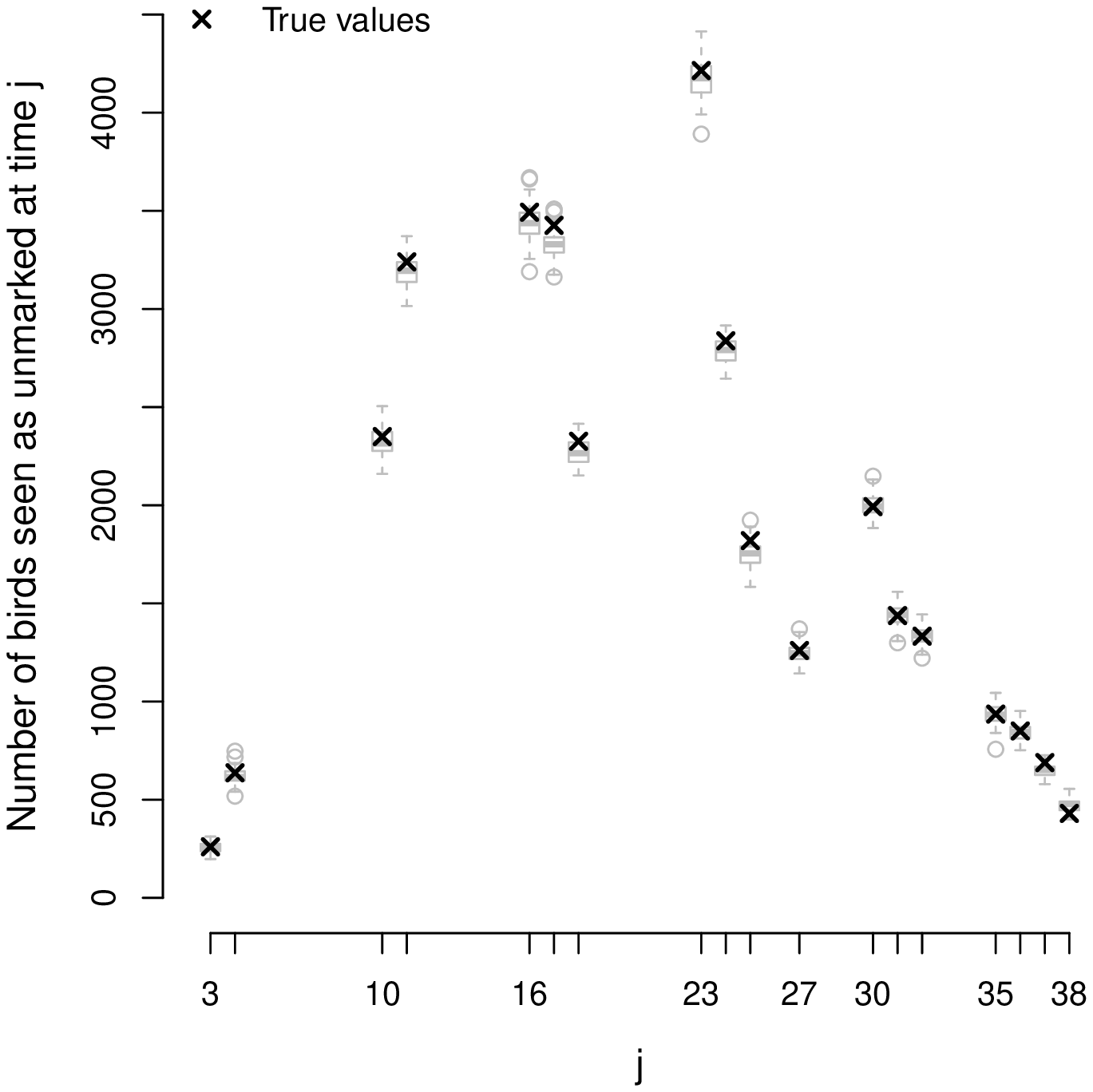}
                            \caption{}
    \end{subfigure}\\
    %\hspace{0.7cm}
    \begin{subfigure}[]{0.45\textwidth}
            \includegraphics[scale=0.35]{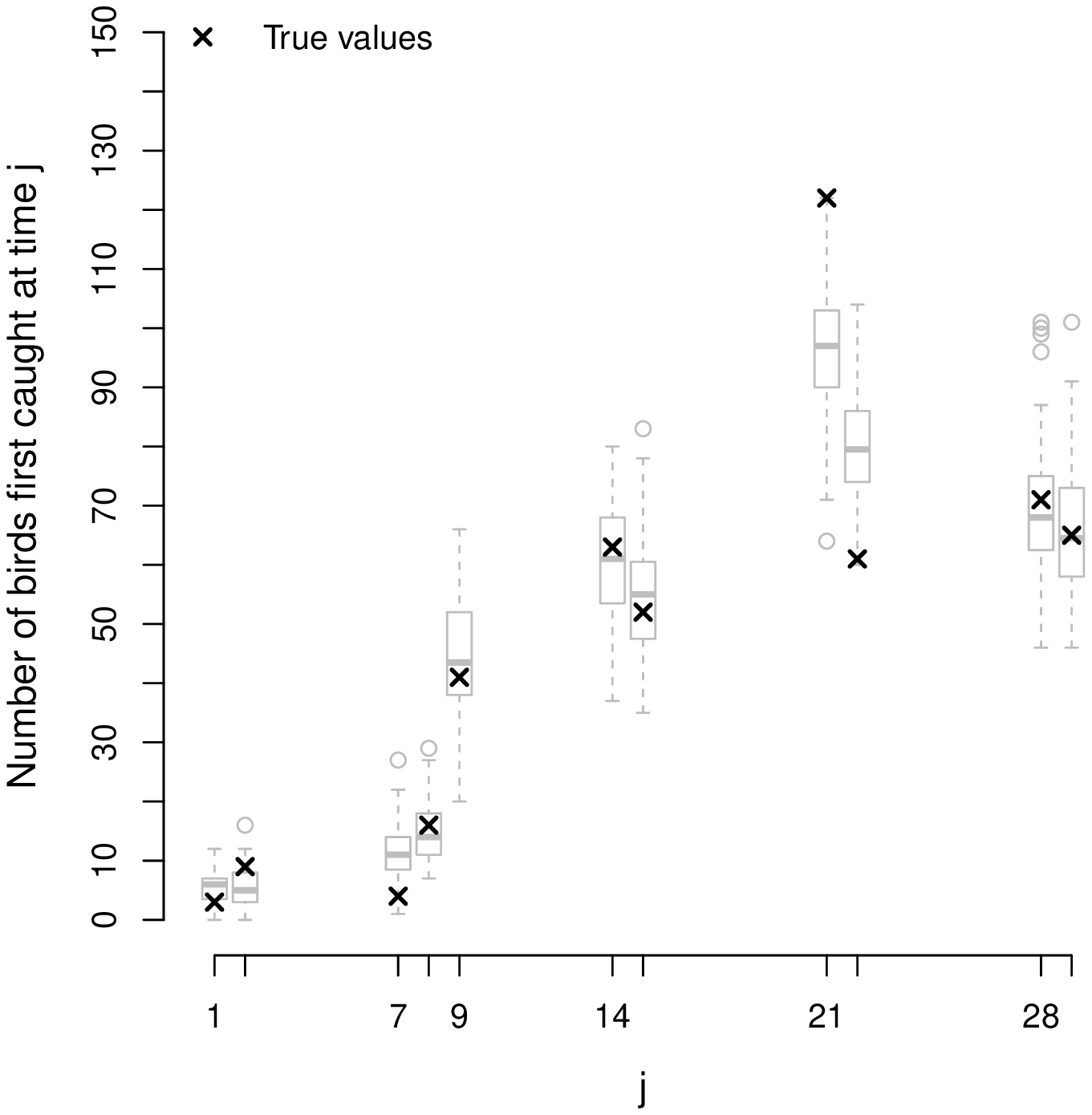}
                            \caption{}
    \end{subfigure}
    \caption{(a): Density of the log-likelihood values obtained at 100 randomly chosen simulation runs for simulated and the true data set.
    (b): Number of birds counted as unmarked at each sample in simulated data sets (boxplots) and the true data set.
    (c): Number of birds first caught at each sample in simulated data sets (boxplots) and the true data set.}
\label{fig:fit}
\end{figure}

\end{document}